\documentclass[prb,twocolumn,superscriptaddress,nofootinbib]{revtex4-2}

\usepackage{amsmath}
\usepackage{graphicx}
\usepackage{xcolor}
\usepackage{hyperref}

\begin{document}

\title{Fluctuation-induced first order phase transitions in Kitaev-like $d^4$ honeycomb magnet}

\date{\today}

\author{Yuchen Fan}
\affiliation{Institute of Physics, Chinese Academy of Sciences, Beijing 100190, China}

\author{Yuan Wan}
\email{yuan.wan@iphy.ac.cn}
\affiliation{Institute of Physics, Chinese Academy of Sciences, Beijing 100190, China}
\affiliation{Songshan Lake Materials Laboratory, Dongguan, Guangdong 523808, China}

\begin{abstract}
We study numerically a bosonic analog of the Kitaev honeycomb model, which is a minimal model for $4d^4/5d^4$ quantum magnets with honeycomb lattice geometry. We construct its phase diagram by a combination of Landau theory analysis and quantum Monte Carlo simulations. In particular, we show that the phase boundaries between the paramagnetic state and magnetically ordered states are generically fluctuation-induced first order phase transitions. Our results are potentially applicable to Ru$^{4+}$- and Ir$^{5+}$-based honeycomb magnets.
\end{abstract}

\maketitle

\section{Introduction}

\begin{figure}
\includegraphics[width = 0.9\columnwidth]{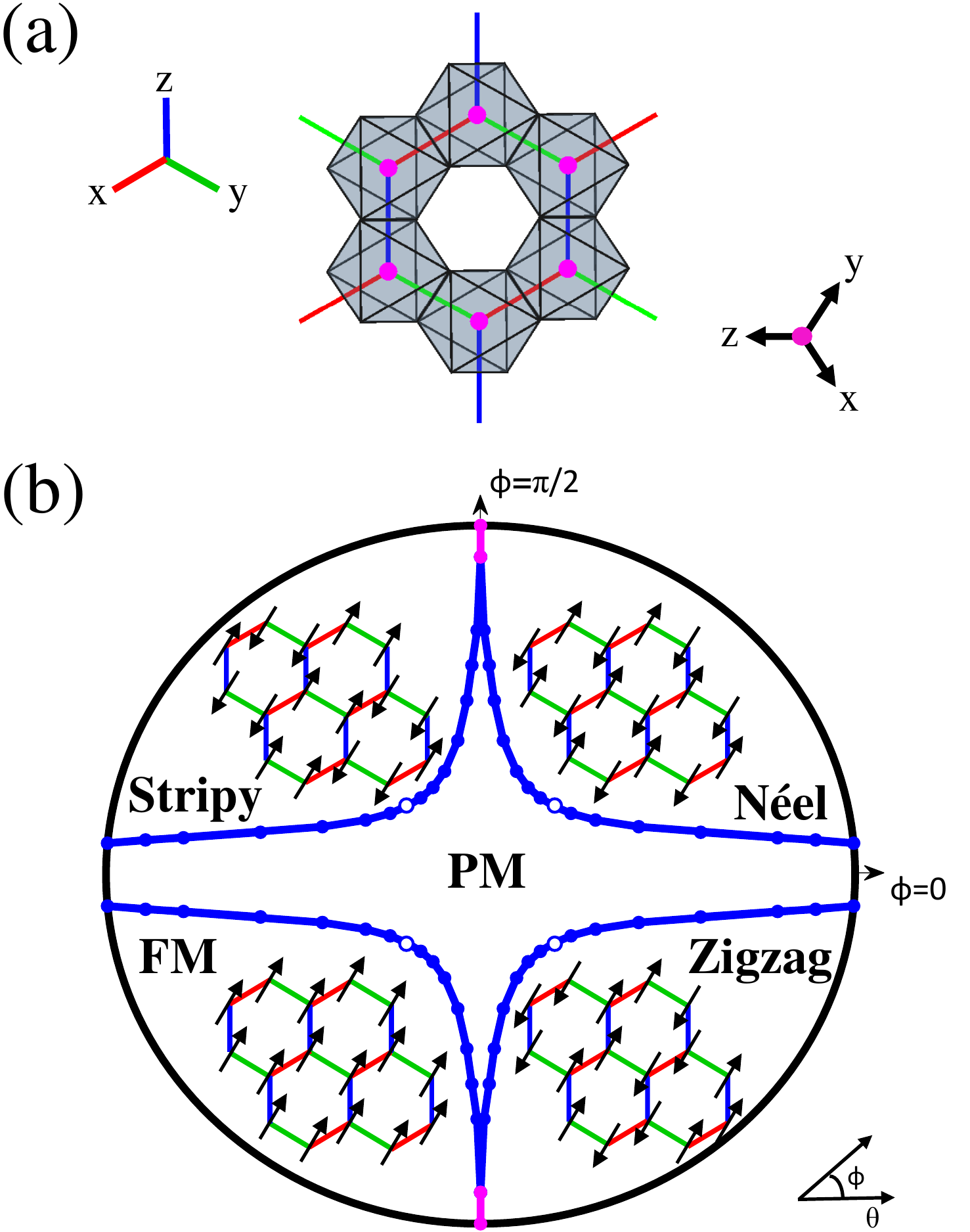}
\caption{(a) In a $d^4$ honeycomb magnet, the $4d^4/5d^4$ transition metal ions form a honeycomb lattice. Each ion is surrounded by an edge-sharing oxygen octahedron. The three translationally inequivalent bonds are dubbed $x$-, $y$-, and $z$-type and respectively colored in red, green, and blue. The top view of the Cartesian axes is also shown. (b) Ground state phase diagram of the bosonic Kitaev model constructed from quantum Monte Carlo. $E_T=\cos\theta$, $K = \sin\theta\cos\phi$, $\widetilde{K} = \sin\theta\sin\phi$. Blue dotted lines represent fluctuation-induced first order phase transitions. Open circles mark continuous phase transitions of the three-dimensional $O(3)$ Heisenberg universality class. Magenta solid lines denote first order boundaries between distinct magnetic orders, which also hosts nematic order. Their end points (magenta dots) are less certain than the other portions of the phase boundaries.}
\label{fig:phase_diagram}
\end{figure}

Magnets made of $4d$ or $5d$ transition metal ions have emerged as a novel platform for exploring quantum magnetism~\cite{Jackeli2009,Chaloupka2010,Chen2010,Khaliullin2013,Witczak2014,Meetei2015,Rau2016,Winter2017,Takayama2021,Trebst2022}. In these systems, the interplay between the spin-orbital coupling and the crystal field environment produces pseudospin degrees of freedom, which can be thought of as entangled textures of both spin and orbital states. Owing to the spin-orbital entanglement, the exchange interaction between the pseudospins can acquire a highly anisotropic form, which, in turn, may lead to a host of unconventional magnetic phases and phase transitions. 

Among the vast family of $4d/5d$ transition metal magnets, $4d^4/5d^4$ honeycomb magnets~\cite{Khaliullin2013,Anisimov2019,Chaloupka2019}, represented by Ag$_3$LiRu$_2$O$_6$~\cite{Kimber2010,Kumar2019}, Na$_2$RuO$_3$~\cite{Mogare2004,Wang2014,Gapontsev2017,Veiga2020}, and potentially SrIr$_2$O$_6$~\cite{Song2021}, stand out as a unique class. Here, the ions form a honeycomb lattice, with each ion being at the center of an edge-sharing oxygen octahedron (Fig.~\ref{fig:phase_diagram}a). The ground state of an isolated ion is a pseudospin singlet $J=0$. The first excited states are the $J=1$ triplet, known as the triplons. The triplons can hop between the neighboring sites and create/annihilate in pairs due to the exchange interactions. Sufficiently strong exchange interactions may overcome the triplon energy gap and trigger the condensation of the triplons~\cite{Giamarchi2008}, thereby inducing a magnetic order in a nominally non-magnetic system. 

While similar exchange-induced singlet magnetism is known to exist in rare earth magnets~\cite{Cooper1972}, a defining feature of the $d^4$ honeycomb magnet is its highly anisotropic exchange interaction~\cite{Khaliullin2013}. Owing to the spin orbital entanglement inherent to the triplons, as well as the edge-sharing geometry of the oxygen octahedra, the exchange interaction depends strongly on both the triplon pseudospin and the bond direction. This form of anisotropic exchange interaction is reminiscent of the celebrated Kitaev honeycomb model~\cite{Kitaev2006}, incarnated in $4d^5/5d^5$ honeycomb magnets~\cite{Chaloupka2010}, with the important distinction that the former is a system of $J=1$ triplons and the latter $J=1/2$ pseudospins.

The minimal model that captures the essential physics of the $d^4$ honeycomb magnet is thus a bosonic analog of the Kitaev honeycomb model~\cite{Khaliullin2013,Anisimov2019,Chaloupka2019}, which, for brevity, we refer to as the \emph{bosonic Kitaev model} henceforth. This model is a singlet-triplet model consisting of two terms, an on-site potential energy term and an exchange term. The potential term controls the bare energy cost for creating a triplon ($E_T$), whereas the exchange term describes the bond-dependent coupling between the van Vleck magnetic moments associated with the admixture of the singlet and the triplet. The exchange term features two independent parameters, $K$ and $\widetilde{K}$, that interpolate various limits~\cite{Chaloupka2019}. In particular, in the limit $K\neq 0$, $\widetilde{K}=0$, the exchange interaction is a bond-dependent Ising interaction akin to the Kitaev honeycomb model. By contrast, the exchange interaction is isotropic in the limit $K=\widetilde{K}$.

Previous studies on the bosonic Kitaev model have revealed interesting properties of the model~\cite{Khaliullin2013,Anisimov2019,Chaloupka2019}. In the aforementioned Kitaev limit, the anisotropic exchange interaction frustrates the propagation of triplons and prevents them from condensing even when the bare value of the triplon energy gap $E_T$ is set to zero~\cite{Chaloupka2019}. In other words, the system remains a correlated quantum paramagnet throughout the entire $E_T$ axis. The phase diagram is less clear away from this limit. By using analytical approximations and the exact diagonalization on small clusters~\cite{Anisimov2019,Chaloupka2019}, previous works have sketched the qualitative structure of the phase diagram. Yet, the precise structure of the phase diagram as well as the nature of the phase transitions have not been determined so far. 

In this work, we construct the phase diagram of the bosonic Kitaev model from quantum Monte Carlo simulations. Our main result is the ground state phase diagram shown in Fig.~\ref{fig:phase_diagram}b. The system hosts respectively the N\'{e}el, stripy, ferromagnetic, and zigzag magnetic order in its first, second, third, and fourth quadrants. Furthermore, we find that the transitions from the paramagnetic state to these magnetic ordered states across \emph{generic} positions of the phase boundary are all fluctuation-induced first order transitions. This occurs because the combination of the model's internal symmetry and the lattice symmetry give rise to an effective cubic symmetry for the magnetic order parameters. The transition is in the same universality class as the $O(3)$ Heisenberg model with face-centered cubic anisotropy, which is known to exhibit fluctuation-induced first order transitions~\cite{Aharony1976,Nienhuis1983,Roelofs1993,Cardy1996}.

The rest of this work is organized as follows. We review the essentials of the bosonic Kitaev model in Sec.~\ref{sec:model}. Based on a Landau theory analysis, we argue that the magnetic phase transitions in this system must be fluctuation-driven first order phase transitions. After a brief discerption of the numerical method in Sec.~\ref{sec:method}, we present our data in Sec.~\ref{sec:results}. Finally, we discuss a few open questions in Sec.~\ref{sec:discussion}.

\section{Model \label{sec:model}}

In this section, we discuss the bosonic Kiatev honeycomb model in detail. We give a brief account of the model in Sec.~\ref{sec:hamiltonian}. In Sec.~\ref{sec:symmetries}, we discuss the symmetries of the model Hamiltonian, which is indispensable for understanding the magnetic orders and phase transitions in this system. In particular, the four quadrants of the parameter space spanned by the coupling constants $K$ and $\widetilde{K}$ are related by gauge transformations. As a result, it is sufficient to consider the first quadrant $K\ge 0$ and $\widetilde{K}\ge0$. Finally, in Sec.~\ref{sec:Landau}, we argue that the system may develop N\'{e}el order when $K>0$ and $\widetilde{K}>0$. We construct a Landau theory for the N\'{e}el order and argue that the transition from the paramagnetic state to the N\'{e}el state must be weakly first order owing to the fluctuation-induced first order transition mechanism.

\subsection{Hamiltonian \label{sec:hamiltonian}}

To set the stage, we give a brief review of the bosonic Kiatev honeycomb model. Its basic building blocks are the triplon creation/annihilation operators. Denoting the singlet state by vacuum $|0\rangle$, acting $T^\dagger_{ia}$ on $|0\rangle$ creates a triplon on honeycomb site $i$ with flavor $a$. As each triplon carries total angular momentum $J = 1$, it comes in three possible flavors $a=x,y,z$. Here, $x,y,z$ refer to the three time-reversal invariant basis states of the $J=1$ spin space, which are related to the more familiar $J^z$ basis states by the following unitary transformation: 
\begin{subequations}
\begin{align}
|x\rangle &= \frac{i}{\sqrt{2}} (|J^z=1\rangle - |J^z=-1\rangle);
\\
|y\rangle &= \frac{1}{\sqrt{2}} (|J^z=1\rangle + |J^z=-1\rangle);
\\
|z\rangle &= -i|J^z=0\rangle.
\end{align}
\end{subequations}
As a result, the tripolon operators $T_{ia}$ are invariant under time reversal. Since each site can host at most one triplon, we must enforce the following constraint on the local Hilbert space:
\begin{align}
n_i = \sum_{a}T^\dagger_{ia} T^{\phantom\dagger}_{ia} \le 1,
\end{align}
In other words, the triplons are hard-core bosons. Owing to this constraint, the dimension of the local Hilbert space is $4 = 1\oplus 3$, i.e. a direct sum of a spin singlet and a triplet. 

Admixture between the vacuum (singlet) state $|0\rangle$ and the triplet state $|a\rangle = T^\dagger_a|0\rangle$ produces van Vleck-like magnetic moment in the Cartesian $a$-axis (Fig.~\ref{fig:phase_diagram}a). Up to a numerical constant, the van Vleck moment in the $a$-axis on site $i$ is given by:
\begin{align}
m^V_{ia} = -\frac{i}{2}(T^{\phantom\dagger}_{ia}-T^\dagger_{ia}),
\label{eq:van_vleck}
\end{align}
Since $m^V_{ia}$ transforms as an axial vector under spatial operations, an immediate consequence of Eq.~\eqref{eq:van_vleck} is that $T_{ia}$ transforms as an axial vector as well. Each triplon also carries a magnetic dipole moment, which is given by (up to a numerical constant): 
\begin{align}
m^T_{ia} = -i \sum_{bc}\epsilon^{abc} T^\dagger_{ib} T^{\phantom\dagger}_{ic}.
\end{align}

The Hamiltonian for the bosonic Kiatev model is given by:
\begin{align}
H = \sum_i E_T n_i +\sum_{\langle ij\rangle_a} K O^a_{ij} + \widetilde{K} (O^{\overline{a}}_{ij}+O^{\overline{\overline{a}}}_{ij}),
\label{eq:Hamiltonian}
\end{align} 
The first summation is over all honeycomb sites. $n_i$ is the triplon occupation number defined previously. $E_T>0$ is the energy cost for creating a triplon. The second summation is over all nearest neighboring bonds. It is convenient to denote the three types of translationally inequivalent bonds by $x$, $y$, and $z$ (Fig.~\ref{fig:phase_diagram}a). The subscript in the notation $\langle ij\rangle_a$ indicates the bond is of type $a$. The operator $O^a_{ij}$ is the coupling of the van Vleck moments on site $i$ and $j$ in the $a$-axis. In terms of triplons, $O_{ij}$ describes the hopping, as well as the pair production and annihilation, of triplons with the flavor $a$ on the bond $ij$:
\begin{align}
O^a_{ij} = 4m^V_{ia} m^V_{ja} = T^\dagger_{ia}T^{\phantom\dagger}_{ja} - T^\dagger_{ia}T^{\dagger}_{ja} + H.c. .
\end{align}
$K$ and $\widetilde{K}$ parametrize the magnitude for these couplings. Note the magnitude depends on both the direction of the bond and the triplon flavor.  On an $a$-type bond, the interaction magnitude for the triplons with the matching flavor, namely $O^a_{ij}$, is given by $K$, whereas the magnitude for the complementary flavors $\overline{a}$ and $\overline{\overline{a}}$ is given by $\widetilde{K}$. This anisotropic, bond-dependent interaction is strongly reminiscent of the situation one encounters with the celebrated $S=1/2$ Kiatev honeycomb model, and, therefore, Eq.~\eqref{eq:Hamiltonian} may be viewed as its bosonic analog.

The parameter space of the model Hamiltonian is two-dimensional. Following Ref.~\onlinecite{Chaloupka2019}, we reparametrize the coupling constants as $E_T = \cos\theta$,  $K = \sin\theta\cos\phi$, and $\widetilde{K} = \sin\theta\sin\phi$. We are only interested in the regime $E_T\ge 0$ ($\theta\in [0,\pi/2]$), i.e. the triplet state has higher energy than the singlet state. The regime $E_T<0$, while interesting in itself~\cite{Chaloupka2019}, is not directly realized in the context of the honeycomb $d^4$ magnets. 

\subsection{Symmetries \label{sec:symmetries}}

\begin{figure}
\includegraphics[width=\columnwidth]{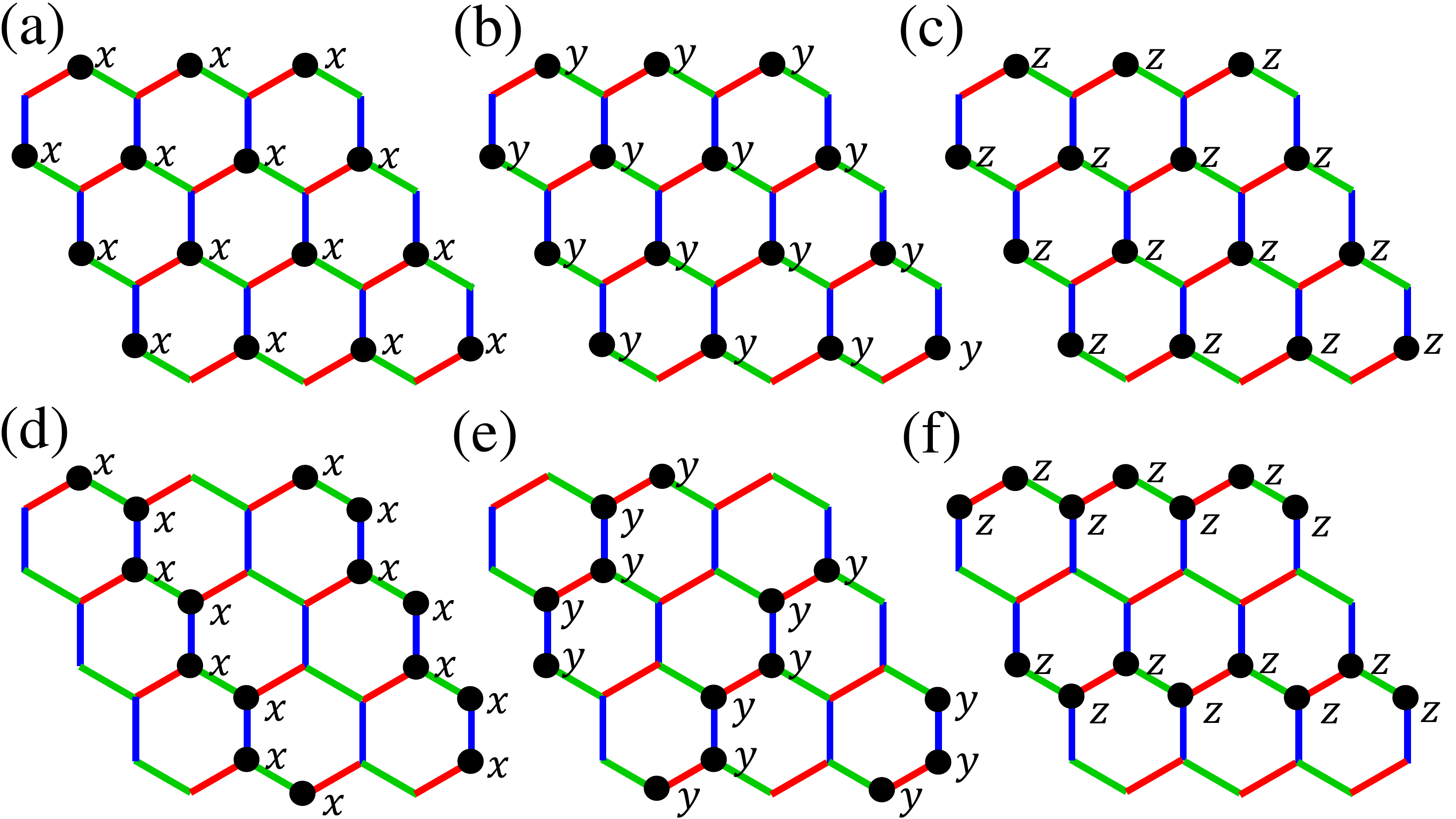}
\caption{(a)(b)(c) The gauge transformation that maps $K\to -K$ and $\widetilde{K}\to -\widetilde{K}$. It comprises of three separate transformations for each flavor of triplons, show in three panels. A site $i$ highlighted with a black dot and marked with the label $a$ indicates that a gauge transformation $T_{ia} \to -T_{ia}$ is performed on this site. (d)(e)(f) The gauge transformation that maps $K\to -K$ while keeps $\widetilde{K}$ invariant. }
\label{fig:gauge}
\end{figure}

The model Eq.~\eqref{eq:Hamiltonian} features a rich set of symmetries, which we shall enumerate in this section. To this end, we recall that each honeycomb site, occupied by a $d^4$ transition metal ion, is decorated by an octahedral oxygen environment. The crystallographic point group of this lattice is $D_{3d}$.

Eq.~\eqref{eq:Hamiltonian} is apparently invariant under all lattice translations as well as the time reversal. We also verify that the model is invariant under all the point group operations accompanied by appropriate transformation of $T_{ia}$ operators. It is sufficient to consider two generators of the point group, namely an $S_6$ ($\pi/6$ rotation with respect to the center of the hexagon plus a mirror reflection with respect to the honeycomb plane), and a $\sigma_d$ reflection whose mirror plane cuts the hexagonal bond in the middle. For the former operation, a simultaneous cyclic permutation of the triplon operator $T_{ix}\to T_{iz}\to T_{iy}\to T_{ix}$ restores the Hamiltonian; for the latter, the transformation: $T_{ix} \to -T_{iy}$, $T_{iy}\to -T_{ix}$, $T_{iz}\to -T_{iz}$, renders the Hamiltonian invariant. 

In addition to the lattice symmetries and the time reversal symmetry,  there is a $Z_2$ internal symmetry associated with \emph{each} triplon flavor: $T_{ia} \to -T_{ia}$. The overall internal symmetry is thus $Z_2^{\times 3}$. We stress that this internal symmetry is a unique feature of the model Eq.~\eqref{eq:Hamiltonian}. Strictly speaking, in a system with strong spin-orbital coupling, the only microscopic symmetries (i.e. symmetries of the Hamiltonian on the lattice scale) are the lattice symmetries and the time-reversal symmetry. Indeed, microscopic considerations show that interaction terms involving three triplon operators, even though small in magnitude, nevertheless appear and weakly break the $Z_2^{\times 3}$ internal symmetry~\cite{Khaliullin2013}. We shall comment further on this point in Sec.~\ref{sec:discussion}

So far our discussion has been for generic points of the parameter space. Yet more symmetries emerge at several corners of the parameter space.  First of all, in the Kitaev limit ($K\neq 0$ and $\widetilde{K} = 0$), the model has an extensive number ($\propto L^2$, where $L$ is the linear dimension of the lattice) of $Z_2$ symmetries, which prevent the system from developing magnetic order even when $E_T=0$~\cite{Chaloupka2019}. Secondly, in the isotropic limit ($K=\widetilde{K}$), the model possesses an $O(3)$ internal symmetry akin to a rotor model. Finally, in the compass limit ($K=0$ and $\widetilde{K}\neq 0$), the model has a sub-extensive number ($\propto L$) of $Z_2$ symmetries~\cite{Takayama2021}. This situation is reminiscent of the spin compass model, and hence the nomenclature~\cite{Dorier2005,Wenzel2008,Nussinov2015}. We shall examine this limit in Sec.~\ref{sec:compass}.

Similar to the Klein duality of the Kitaev honeycomb model~\cite{Chaloupka2010,Chaloupka2016}, the four quadrants of the parameter space are related by gauge transformations. To see this, we observe that we may perform gauge transformations individually on each flavor of the triplons, $T_{i\alpha} \to \eta_{i\alpha} T_{i\alpha}$, where $\eta_{i\alpha} = \pm 1$ is a $Z_2$ phase. The gauge transformation shown in the top row of Fig.~\ref{fig:gauge} maps $K$ to $-K$ and $\widetilde{K}$ to $-\widetilde{K}$, whereas the transformation shown in the bottom row of Fig.~\ref{fig:gauge} maps $K\to -K$ but keeps $\widetilde{K}$ invariant. Therefore, in the ensuing discussion, we focus on the first quadrant of the parameter space: $K\ge 0$ and $\widetilde{K}\ge 0$, or equivalently $\phi\in[0,\pi/2]$.

\subsection{Landau theory \label{sec:Landau}}

The main interest of this work is the potential magnetic order hosted by the bosonic Kitaev model, and the nature of the magnetic phase transition associated with it. In Sec.~\ref{sec:symmetries}, we have pointed out that it is sufficient to consider the first quadrant of the parameter space, $K\ge 0$ and $\widetilde{K}\ge 0$. We now argue that the natural candidate magnetic order is the N\'{e}el order in this quadrant. 

To this end, we note that the interaction terms are in the form of $m^a_i m^a_j$, which favors the N\'{e}el magnetic order when both $K$ and $\widetilde{K}$ are positive. The remaining question is the direction of the N\'{e}el vector. Now, owing to the hard-core constraint of the triplons and the form of the coupling, we expect that it is energetically unfavorable for all three flavors of triplons to condense simultaneously. Said differently, it is more likely for the N\'{e}el vector to point in the $\pm x$, $\pm y$, or $\pm z$ axis (the $[100]$ direction), as opposed to, for instance, the $[111]$ direction~\footnote{It is possible to approximate the model Eq.~\eqref{eq:Hamiltonian} as a quantum rotor model, where the first term of the Hamiltonian plays the role of rotor kinetic energy, whereas the second term the rotor coupling. When the interaction dominates over the kinetic energy, one may perform a linear spin wave theory calculation and show that the $[100]$ direction is selected by the order by quantum disorder mechanism.}.

We now analyze the N\'{e}el order by using the Landau theory. We describe the N\'{e}el order by the three Cartesian components of the N\'{e}el vector $\mathbf{N}$, namely $N_x$, $N_y$, and $N_z$. They are invariant under lattice translations, and odd under time reversal. The $S_6$ operation induces the transformation $N_x\to -N_y$, $N_y\to -N_z$, and $N_z\to -N_x$. Meanwhile, under the $\sigma_d$ operation, $N_x\to N_y$, $N_y\to N_x$, $N_z\to N_z$. Furthermore, the internal $Z_2^{\times 3}$ symmetry operation induces a transformation $N_a\to -N_a$, for each and every $a$. Up to quartic order, these symmetries constrains the Landau free energy to the following form:
\begin{align}
F = \alpha \mathbf{N}^2 + \beta (\mathbf{N}^2)^2 + \gamma (N^4_x+N^4_y+N^4_z).
\label{eq:Landau}
\end{align}
Here, the sign change in $\alpha$ drives the N\'{e}el phase transition. $\beta>0$. Crucially, $\gamma<0$, which reflects the fact that the three components of the N\'{e}el vector compete rather than cooperate with each other.

We recognize that the Landau free energy has the same form as the $O(3)$ model with cubic anisotropy. In two and three dimensions, it has been shown that the $O(3)$ model exhibits fluctuation-driven first order phase transitions \emph{provided} that the sign of the cubic anisotropy term $\gamma$ is \emph{negative}, i.e. the cubic anisotropy is of the face-centered type~\cite{Aharony1976,Nienhuis1983,Roelofs1993,Cardy1996}. Utilizing this fact, we expect that the thermal and quantum phase transitions from the paramagnetic state to the N\'{e}el state to be both weakly first order. We note that, at this stage, we cannot completely rule out other possibility such as the two-step transition.

We close this section by discussing the other quadrants of the parameter space. Owing to the gauge transformations (Sec.~\ref{sec:symmetries}), the magnetic order in the other three quadrants of the parameter space are respectively the stripy order, the ferromagnetic order, and the zigzag order. The associated magnetic phase transitions are all weakly first order.  

\section{Method \label{sec:method}}

We perform a quantum Monte Carlo simulation of the bosonic Kitaev model Eq.~\eqref{eq:Hamiltonian} based on the stochastic series expansion algorithm~\cite{Syljuasen2002,Sandvik2010}. We operate in the occupation number basis of the triplons; the Hamiltonian is free from sign problems in this basis. We employ a lattice of $L\times L$ primitive unit cells with the periodic boundary condition enforced in both lattice directions. The maximal linear dimension $L=48$. The lowest simulated temperature $T=0.01$. We use $>40$ parallel Markov chains to obtain statistically independent measurements of observables. Each chain contains $>10^4$ Monte Carlo steps (MCS) for thermalization and $>2\times 10^3$ MCS for sample accumulation. 

We employ the following Monte Carlo observables to detect the magnetic phases and phase transitions. We estimate the specific heat per site $C$ by approximate it as the energy difference:
\begin{align}
C \approx \frac{\langle H\rangle (T+\Delta T)-\langle H\rangle (T)}{2L^2 \Delta T}.
\end{align} 
where $2L^2$ is the total number of sites.

We detect the potential magnetic ordering by using the structure factor:
\begin{align}
S(\mathbf{q}) = \frac{1}{2L^2}\sum_{a} \sum_{i,j} \eta_i\eta_j \langle m^V_{ia} m^V_{ja} \rangle e^{i\mathbf{q} \cdot (\mathbf{r}_i-\mathbf{r}_j)}.
\end{align}
Here, $\mathbf{q}$ is the wave vector. $\mathbf{r}_i$ is the real space position of the honeycomb site $i$. $\eta_i = 1(-1)$ if the site $i$ is in the A(B) sublattice. The staggered phase factor $\eta_i$ is convenient for detecting the N\'{e}el order in the honeycomb lattice.  Specifically, the order parameter for the N\'{e}el order coincides with $S(\mathbf{q}=0)$:
\begin{align}
\langle N^2 \rangle = \sum_a \left\langle \left(\frac{1}{2L^2}\sum_{i}\eta_i m^V_{ia} \right)^2 \right\rangle  = \frac{S(\mathbf{q}=0)}{2L^2}.
\end{align}
We do not observe any Bragg peaks from $S(\mathbf{q})$ other then those indicating the N\'{e}el order.

The order parameter $N^2$ is sensitive to the time reversal symmetry breaking but silent for the breaking of the lattice rotational symmetry. To this end, we define an quantity:
\begin{align}
\psi_i = n_{ix} + n_{iy} e^{i\frac{2\pi}{3}} + n_{iz} e^{-i\frac{2\pi}{3}},
\end{align}
where $n_{ia}$ is the triplon occupation number of flavor $a$ on site $i$. In particular, $\psi_i = 1$ if the site is occupied by the $x$ triplon, $\exp(i2\pi/3)$ if by the $y$ triplon, and $\exp(-2i\pi/3)$ if by the $z$ triplon. We define the structure factor associated with the order parameter $\psi$:
\begin{align}
C(\mathbf{q}) = \frac{1}{2L^2}\sum_{a} \sum_{i,j} \langle \psi_i \psi^\ast_j \rangle e^{i\mathbf{q} \cdot (\mathbf{r}_i-\mathbf{r}_j)}.
\end{align}
The order parameter for the three-fold lattice symmetry breaking is given by:
\begin{align}
\langle |\psi|^2\rangle = \left \langle \left( \frac{1}{2L^2}\sum_i \psi_i \right)^2 \right\rangle = \frac{C(\mathbf{q}=0)}{2L^2}.
\end{align}
We do not observe any Bragg peaks from $C(\mathbf{q})$ other than those equivalent to the $\mathbf{q}=0$ point.

Finally, we extract the \emph{effective} correlation lengths from the structure factors as follows~\cite{Sandvik2010}:
\begin{align}
\xi _{N}=\frac{L}{2\pi}\sqrt{\frac{S(0)}{S(\frac{2\pi}{L}\mathbf{g}_1)}-1},
\label{eq:xi_def}
\end{align} 
where $\mathbf{g}_1$ is one of the reciprocal wave vectors of the lattice. $(2\pi/L) \mathbf{g}_1$ is thus one of the smallest non-zero wave vector one may resolve with a lattice consisting of $L\times L$ primitive unit cells.  This definition is motivated by assuming the Lorentz-like form of the structure factors near the ordered wave vector $S(\mathbf{q})\approx S(0)/(1+\mathbf{q}^2\xi^2_N)$. We stress that $\xi_N$ has the same scaling behavior as the true correlation length but do not necessarily the same numerical value. The effective correlation length $\xi_\psi$ may be extracted from $C(\mathbf{q})$ in the same vein.

\section{Results \label{sec:results}}

In this section, we map out the phase diagram of the bosonic Kitaev model (Eq.~\eqref{eq:Hamiltonian}) in the first quadrant of the parameter space and clarify the nature of the phase transitions. In Sec.~\ref{sec:thermal}, we provide numerical evidence for the weakly first order thermal transition into the N\'{e}el state for a representative set of model parameters. In Sec.~\ref{sec:quantum},  we show that the quantum phase transition into the N\'{e}el state is also weakly first order, and demarcate the phase boundary of the N\'{e}el state. In Sec.~\ref{sec:compass}, we briefly discuss the physics unique to the compass limit ($K=0$, $K'\neq 0$).  

\subsection{Thermal phase transition into N\'{e}el order \label{sec:thermal}}

\begin{figure}
\includegraphics[width=\columnwidth]{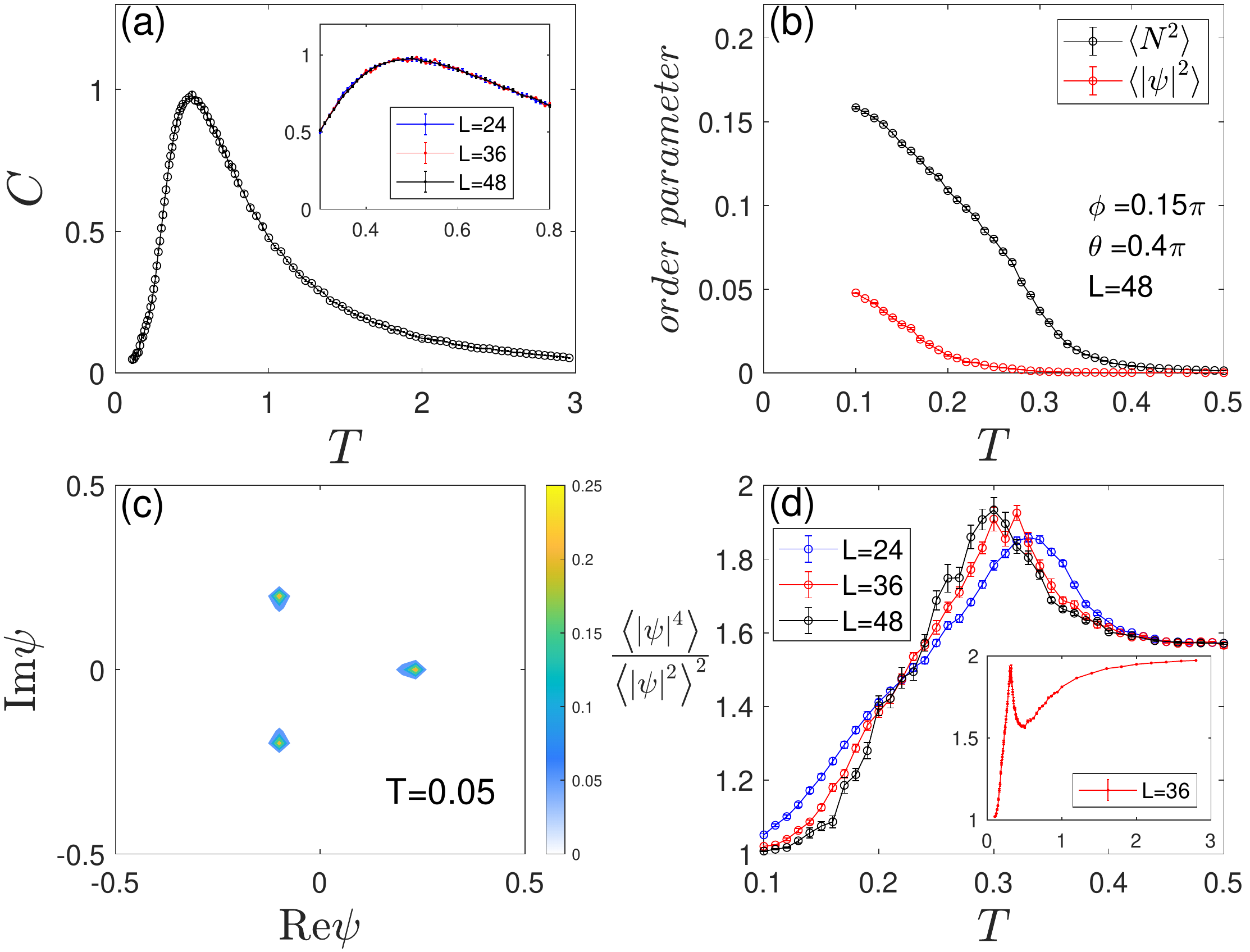}
\caption{The temperature driven phase transition into the N\'{e}el ordered state for the representative model parameters $\phi= 0.15 \pi$, $\theta = 0.4\pi$. The system size $L=48$ unless stated otherwise. (a) The specific heat per site $C$ as a function of temperature $T$. Inset shows the system size dependence. (b) The order parameters $N^2$ and $|\psi|^2$ as a function of $T$. (c) The histogram of the order parameter $\psi$ in the complex plane at $T=0.05$. (d) The binder ratio for the order parameter $\psi$ as a function of $T$ for various system sizes. The inset shows its behavior over a wider temperature window for $L=36$.}
\label{fig:thermal}
\end{figure}

We set the model parameters $\phi = 0.15\pi$ and $\theta = 0.4\pi$, and examine the behavior of the model as it cools down from the high temperature paramagnetic state (Fig.~\ref{fig:thermal}). We find the specific heat shows a broad peak near $T\approx 0.5$ (Fig.~\ref{fig:thermal}a). However, this peak does not reflect phase transitions in that it shows no discenible system size dependence (Fig.~\ref{fig:thermal}a, inset). Indeed, as we shall see momentarily, the true phase transition occurs at a much lower temperature scale. We think this specific heat peak is associated with the temperature scale below which the short-range spin correlations set in, typical for frustrated magnets. 

While the specific heat data do not show any signature of long-range order, the N\'{e}el order parameter $N^2$ shows rapid development beneath its onset temperature (Fig.~\ref{fig:thermal}b, black dots). This suggests that the system hosts a N\'{e}el order, in agreement with our argument in Sec.~\ref{sec:Landau}. Furthermore, the complex modulus of the order parameter $\psi$, which is sensitive to the breaking of spatial rotational symmetry, also sets in at lower temperature (Fig.~\ref{fig:thermal}b, red dots). The breaking of the lattice rotational symmetry is consistent with our picture for the N\'{e}el order.

We ascertain the N\'{e}el vector indeed is aligned in the $x,y,z$ axis by examining the complex phase of the order parameter $\psi$. Fig.~\ref{fig:thermal}c shows the histogram of $\psi$ in the complex plane at $T=0.05$. We observe three sharp peaks in the complex plane at phase angles $\mathrm{arg}\psi = 0$ and $\pm 2\pi/3$. Each angle corresponds to a condensate formed by a single flavor of triplons. In terms of the N\'{e}el vector, this result indicates that it is aligned in $\pm x$, $\pm y$, or $\pm z$ direction, in full agreement with our expectation.

The first piece of evidence for first order phase transition comes from the Binder ratio of the order parameter $\psi$ (Fig.~\ref{fig:thermal}d). For all three system sizes, the Binder ratio shows non-monotonic dependence on temperature (see inset for its behavior over a wider temperature range), a tell-tale sign of first order transitions~\cite{Binder1984}. This non-monotonic behavior becomes more prominent as the system size increases, while the peak position shifts toward low temperature. Meanwhile, we find the Binder ratio crosses approximately at $T\approx 0.22$, suggesting that the system size $L$ may be smaller than the true correlation length near the transition, thereby hinting toward a weakly first order phase transition.

\begin{figure}
\includegraphics[width=\columnwidth]{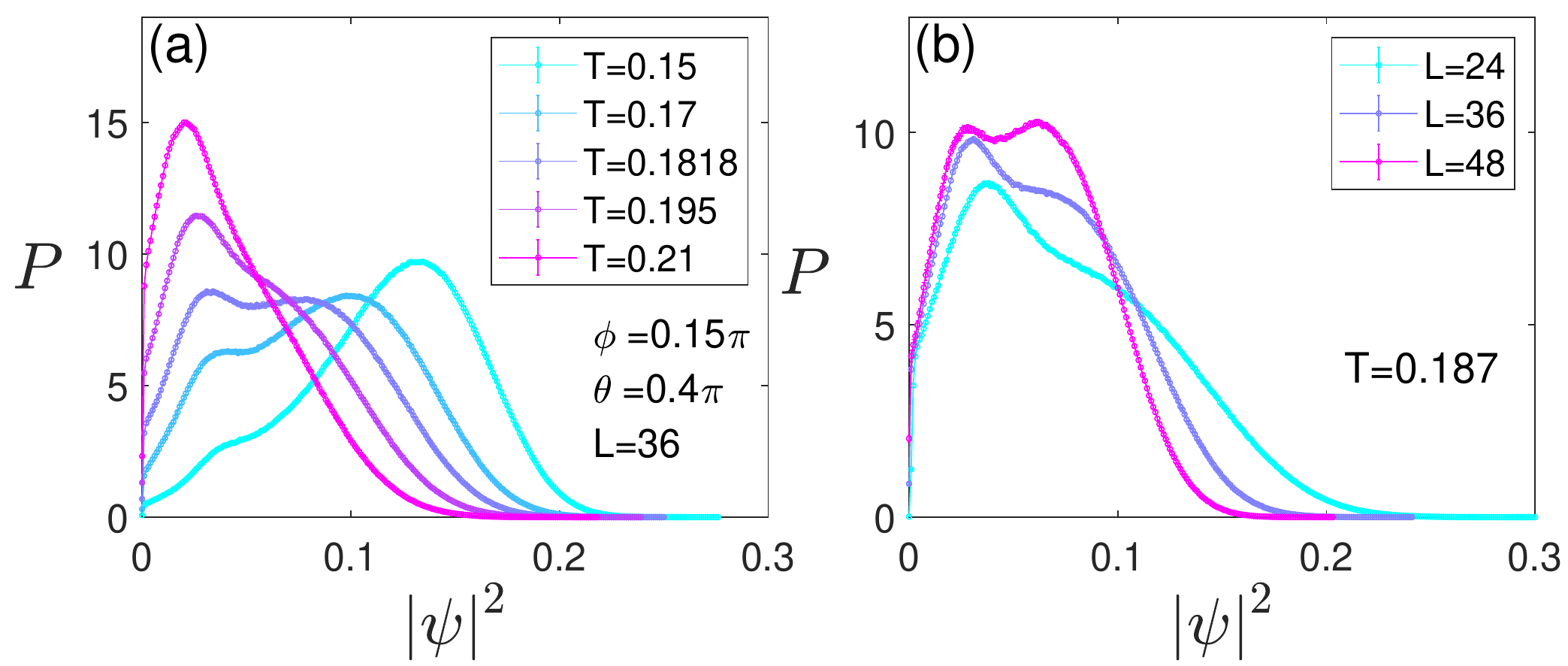}
\caption{Histogram of the complex modulus of the order parameter $\psi$. (a) The temperature-driven evolution of the histogram. The system size $L=36$ (b) The system size dependence of the histogram at fixed temperature $T=0.1870$.}
\label{fig:histo}
\end{figure}

The direct evidence for the first order phase transition lies in the histogram of the complex modulus of the order parameter $\psi$. At both high temperature and low temperature, the histogram shows only a single peak, reflecting a single phase. Importantly, the histogram shows clear double peak structure in a relatively narrow temperature window, between $T=0.15$ and $T=0.21$ for system size $L=36$ (Fig.~\ref{fig:histo}a). This double peak structure is the hallmark of phase coexistence in a first order phase transition. We also examine the finite size effect on the histogram (Fig.~\ref{fig:histo}b). At a given temperature,  the double peak structure is only discernible for sufficiently large system size. This implies that the correlation length near the phase transition may be large, which points toward a weakly first order phase transition as well. 

Interestingly, conventional choices of variables for spotting the first order transitions, such as the histogram of the energy or the triplon density, show single peak over the entire temperature window up to $L=36$ (not shown), which makes the first order transition in this system easy to miss. We think the energy density difference (or the triplon density) between the two phases must be quite small. As a result, observing the double peak structure for these variables requires large system size and high quality data. 

\begin{figure}
\includegraphics[width=\columnwidth]{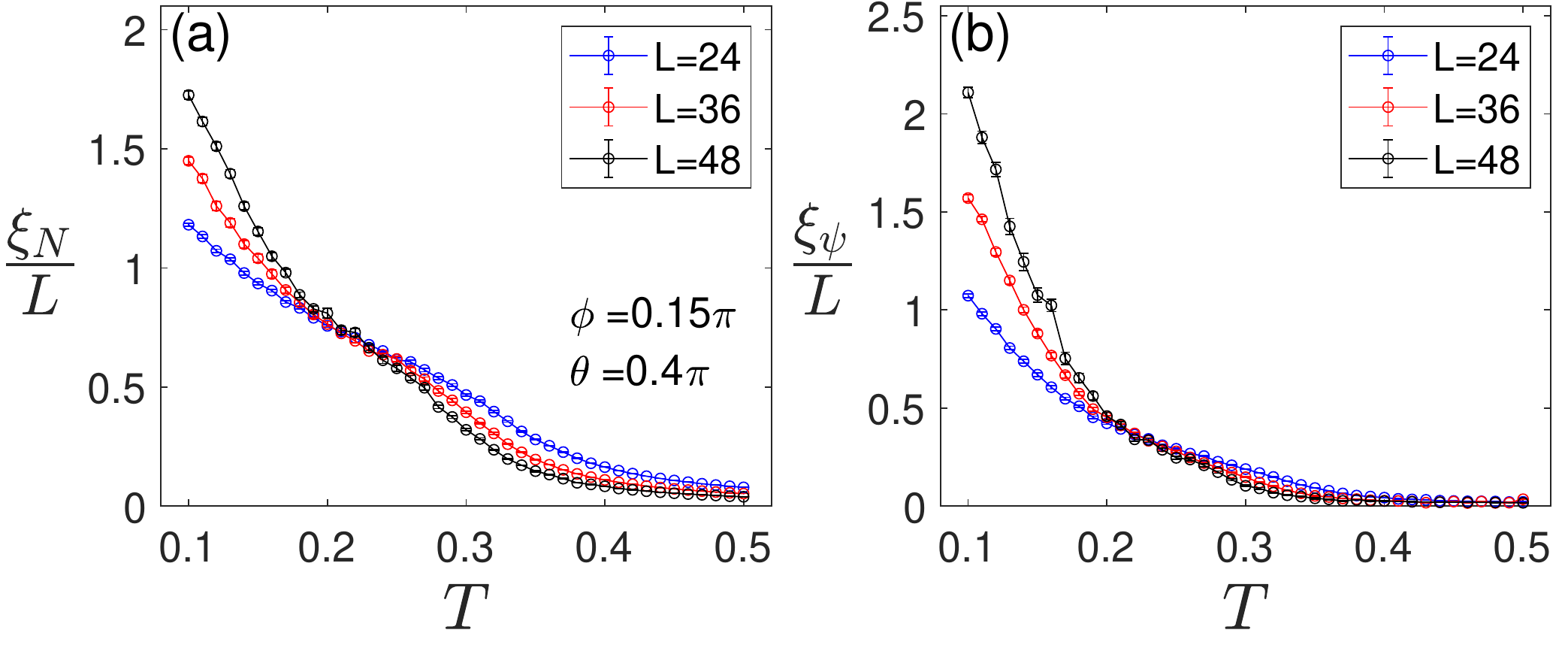}
\caption{Effective correlation length as a function of temperature for (a) the order parameter $N^2$ and (b) the order parameter $\psi$.}
\label{fig:xi}
\end{figure}

Fig.~\ref{fig:xi} shows the effective correlation length associated with the order parameter $N^2$ and $\psi$ . For three different system sizes, we find they cross approximately at $T\approx 0.22$ ($\xi_N$) and $T\approx 0.21$ ($\xi_\psi$), respectively. This is also approximately the temperature at which the Binder ratio crosses (Fig.~\ref{fig:thermal}d). We attribute this approximate crossing to the long correlation length near the weakly first order phase transition --- as long as the system size $L$ does not significantly exceed the correlation length, the transition behaves effectively as a second order phase transition. Note the value of $\xi_{N,\psi}/L$ exceed $1$ because the so-defined $\xi_{N,\psi}$ (Eq.~\ref{eq:xi_def}) are \emph{effective} correlation lengths that share the same scaling behavior as the true correlation length under renormalization group transformations.

\subsection{Quantum phase transition \label{sec:quantum}}

\begin{figure}
\includegraphics[width=\columnwidth]{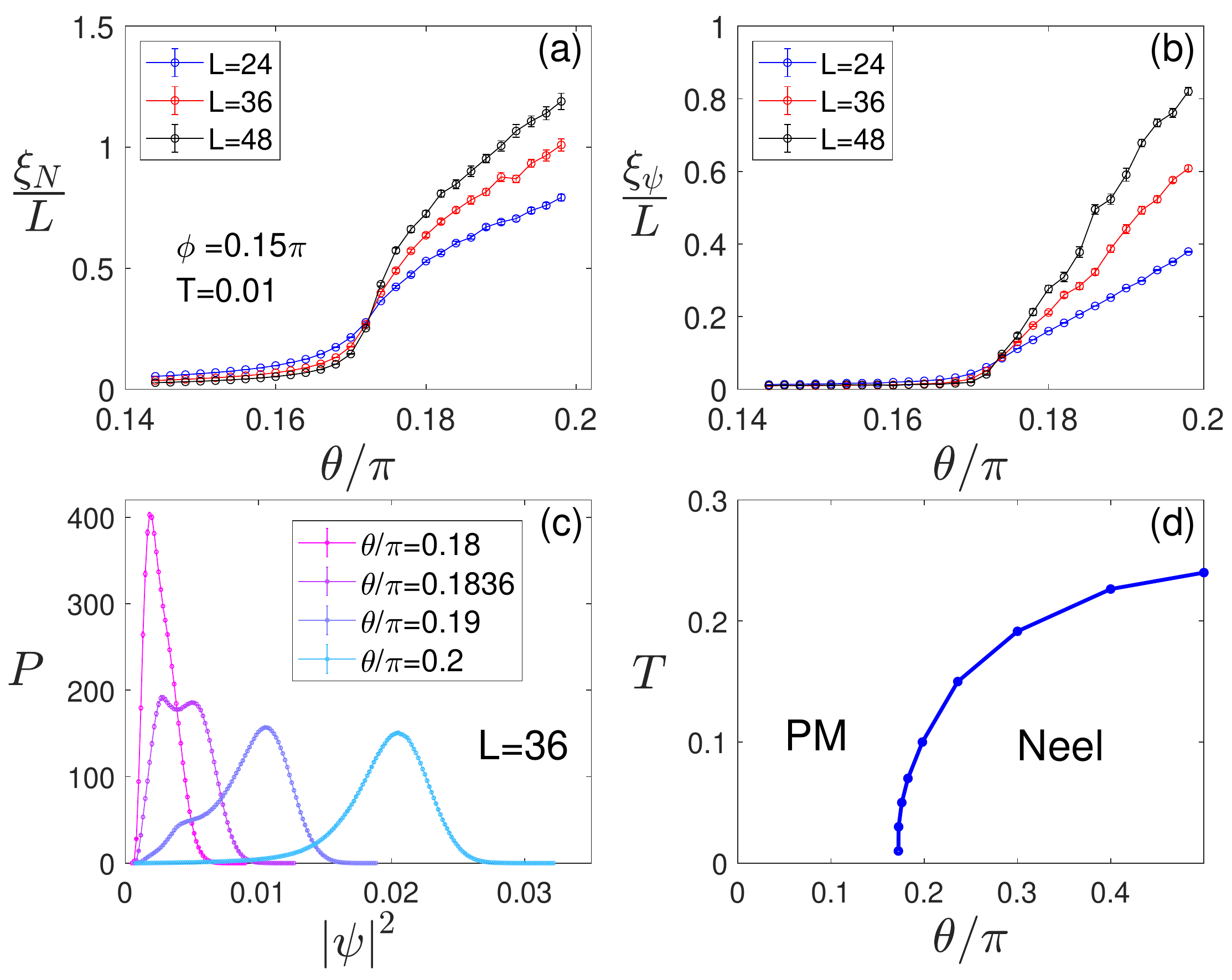}
\caption{Quantum phase transition between the paramagnetic state and the N\'{e}el ordered state driven by the model parameter $\theta$. $\phi=0.15\pi$. (a) The effective correlation length $\xi_N$ as a function of $\theta$ for various system sizes at the lowest simulated temperature $T=0.01$. (b) Same as (a) but for the effective correlation length $\xi_\psi$. (c) The evolution of the histogram for the complex modulus of the order parameter $\psi$ for various values of $\theta$. (d) The phase diagram of the bosonic Kitaev model in the $T$-$\theta$ plane.}
\label{fig:quantum}
\end{figure}

Having established the weakly first order thermal transition into the N\'{e}el order, we now turn to the quantum transition driven by the model parameter $\theta$. As $\theta$ increases from 0 to $\pi/2$, the energy gap for the triplon gradually closes, beyond which point the N\'{e}el order develops. We expect this transition will be weakly first order as well based on the argument given in Sec.~\ref{sec:Landau}.

We examine the $\theta$-driven quantum phase transitions by setting the temperature $T=0.01$. At this temperature, we are effectively probing the ground state physics given the system size used in the simulation ($L\le 48$). We do not observe significant changes in our data if we further decrease the temperature.

The numerical data are consistent with our expectation. The effective correlation length $\xi_N$ and $\xi_\psi$ both shows approximate crossings at $\theta\approx 0.17\pi$ (Fig.~\ref{fig:quantum}a \& b). Meanwhile, the histogram of the complex modulus of the order parameter $\psi$ shows the idiosyncratic double peak structure as $\theta$ approaches the quantum phase transition. Taken together, these data show that the transition is of first order, and that the correlation length at the transition is still large or comparable with the system size used in the simulation. 

Using the crossing point of the effective correlation length $\xi_N$ for two system sizes $L=12$ and $L=24$, we may approximately determine the phase boundary for the N\'{e}el order. Fig.~\ref{fig:quantum}d shows the phase diagram of the bosonic Kitaev model on the $T$-$\theta$ plane at a representative value of $\phi = 0.15\pi$. The paramagnetic state and the N\'{e}el ordered state are separated by a line of weakly first order transitions. We observe that the phase boundary is almost vertical at the lower temperature end,  from which we infer that we are effectively probing the ground state physics at $T=0.01$ for system sizes used in the simulation. We verify that the behavior of the model is qualitatively the same for $\phi = 0.35\pi$ (not shown), another representative value of $\phi$ but with $\widetilde{K}>K$.

Finally, using the same methodology, we may approximately determine the boundary for the N\'{e}el ordered ground state. This yields the phase diagram shown in Fig.~\ref{fig:phase_diagram}b, which is obtained by setting $T=0.01$. A portion of the phase boundary qualitatively matches what has been inferred from the exact diagonalization on small clusters~\cite{Chaloupka2019}~\footnote{J. Chaloupka, private communications.}. Using the gauge transformation (Sec.~\ref{sec:symmetries}), we can obtain the phase diagram for the other three quadrants of the parameter space. 
 
\subsection{Significant others \label{sec:compass}}

\begin{figure}
\includegraphics[width = \columnwidth]{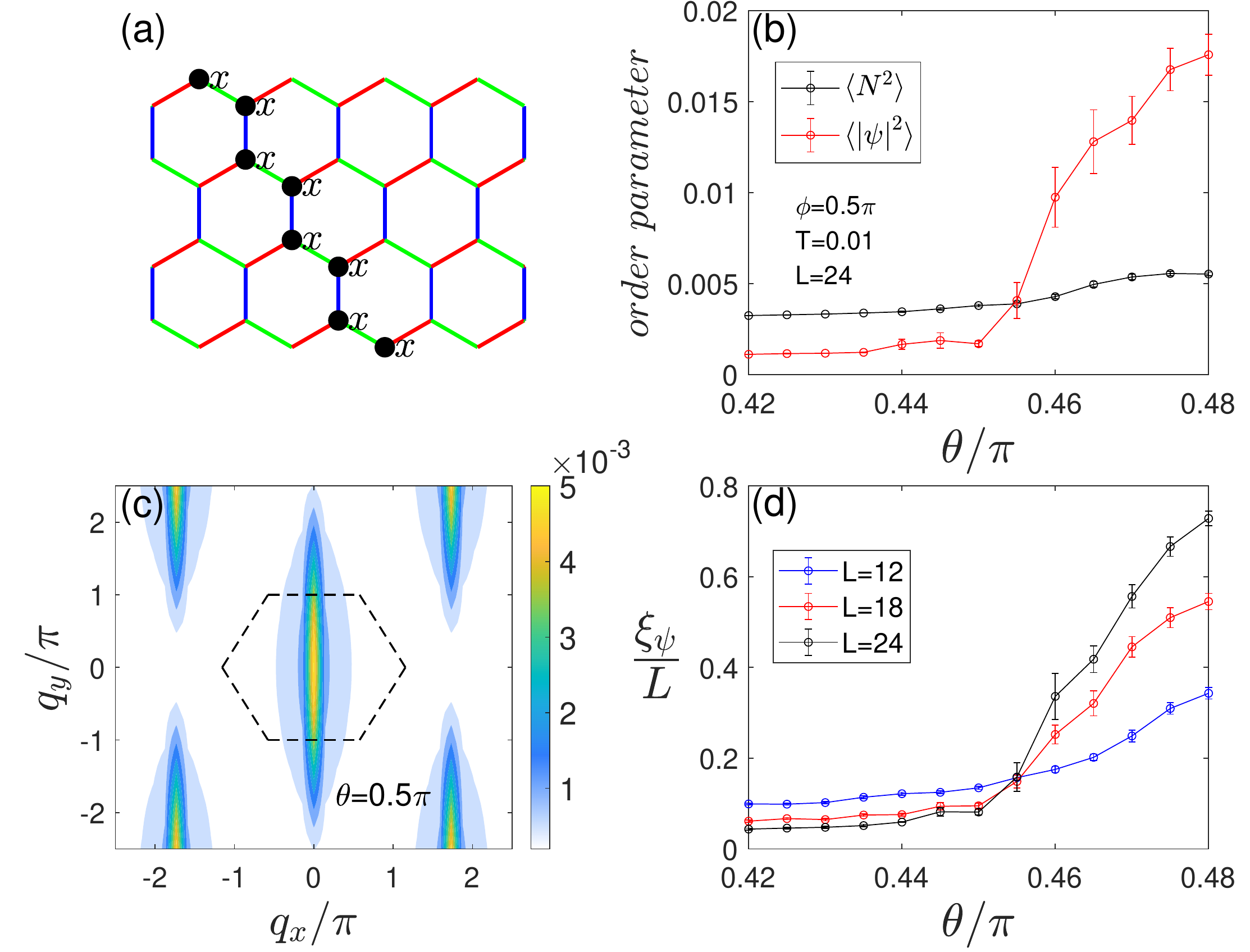}
\caption{(a) A $Z_2$ symmetry operation for the bosonic Kitaev model in the compass limit ($K=0$, $\widetilde{K}\neq 0$). In this limit, the model possesses an sub-extensive ($\sim O(L)$) number of such symmetries. (b) Order parameters $\langle N^2\rangle$ and $\langle |\psi|^2\rangle$ as functions of $\theta$ at $\phi = 0.5\pi$ and $T=0.01$. (c) Spin structure factor $S^{zz}(\mathbf{q})$ at $\theta=0.5\pi$. Dashed line marks the first Brillouin zone boundary. (d) Effective correlation length $\xi_\psi$ as a function of $\theta$.}
\label{fig:compass}
\end{figure}

As we have discussed in Sec.~\ref{sec:symmetries}, at several special corners of the phase diagram, the model possess more internal symmetries in addition to the $Z_2^{\times 3}$ symmetry. As a result, the nature of the magnetic order and the phase transitions might be different from what we have established at generic points of the phase space.

In the Kitaev limit ($\widetilde{K}=0$, or $\phi=0$), the model possesses an extensive number of $Z_2$ symmetries, which prevent the system from developing magnetic order even when $E_T=0$ (equivalently $\theta=\pi/2$)~\cite{Chaloupka2019}. The system is in a correlated, quantum paramagnetic state. From the phase diagram (Fig.~\ref{fig:phase_diagram}b), we see that the quantum paramagnetic state extends to finite value of $\widetilde{K}$ (or $\phi$) as well. 

In the isotropic limit ($K=\widetilde{K}$, or $\phi=\pi/4$), the model possess an internal $O(3)$ symmetry associated with the continuous rotation of the three $T_a$ operators. As a result, the quantum phase transition from the paramagnetic state to the N\'{e}el state along the line $\phi=\pi/4$ is continuous and of the $O(3)$ universality class. Furthermore, the N\'{e}el order does not survive at finite temperature thanks to the Mermin-Wagner theorem.

Less analyzed is the compass limit ($K=0$, or $\phi=\pi/2$). This line is special in that it is the phase boundary between the N\'{e}el order and the stripy order. We now show that the model possesses a sub-extensive number of $Z_2$ symmetries in this limit~\cite{Takayama2021}. To this end, we consider a zigzag path connected by $y$-type and $z$-type bonds shown in Fig.~\ref{fig:compass}a. On this path, the triplons of the flavor $x$ are kinetically \emph{disconnected} from the rest of the lattice. As a result, flipping the sign of the $T_{ix}$ operators on this path does not change the form of the Hamiltonian, and therefore constitutes a $Z_2$ symmetry. Similar symmetric transformations can be constructed for $T_{iy}$ and $T_{iz}$ as well. By construction, the number of such $Z_2$ symmetries is proportional to the linear dimension of the system $L$.

The sub-extensive number of $Z_2$ symmetries prevent the system from developing the N\'{e}el order or zigzag order. Instead, we expect the system to develop a spin nematic order, where the van Vleck moments are strongly correlated along two types of the bonds but uncorrelated along the third. This order breaks the spatial rotational symmetry but preserves the time reversal symmetry. We expect that the transition from the paramagnetic state to this nematic state is of the three-dimensional 3-state Potts type. The situation is analogous to the spin compass model on square lattice~\cite{Dorier2005,Wenzel2008,Nussinov2015}.

Our numerical data are consistent these ideas. As $\theta$ increases toward $\pi/2$, we observe a rapid onset behavior in the order parameter $\langle |\psi|^2\rangle$ (Fig.~\ref{fig:compass}b). However, the order parameter $\langle N^2\rangle$ barely changes. Therefore, the system enters a phase where the lattice rotation symmetry is broken but not the time-reversal symmetry, consistent with the aforementioned nematic order. A direct evidence for the nematic order comes from the spin structure factor for the $z$-components, $S^{zz}(\mathbf{q})$. It shows that the $z$-components are strongly correlated along the horizontal direction but uncorrelated along the vertical direction (Fig.~\ref{fig:compass}c). The structure factors for the other two components $S^{xx}(\mathbf{q})$ and $S^{yy}(\mathbf{q})$ are related to $S^{zz}(\mathbf{q})$ by $2\pi/3$ rotations (not shown). In short, the simulation shows clear signatures of the nematic phase as $\theta$ approaches $\pi/2$.
 
On symmetry ground, we expect that the quantum phase transition from the paramagnetic phase to the nematic phase is in the same universality class as the three-dimensional 3-state Potts model. This transition is known to be first order~\cite{Lee1991}. We find the quantum Monte Carlo algorithm becomes significantly less effective in this parameter regime. Therefore, we are yet unable to clearly resolve the double-peak structure in the histogram of $|\psi|^2$ as we have done for the other parts of the phase diagram. Meanwhile, Fig.~\ref{fig:compass}d shows the effective correlation length $\xi_{\psi}$ as a function of $\theta$ for various system sizes. We find they approximate cross at the same $\theta\approx 0.455\pi$. This suggests that the system size $L$ is not large enough to probe the true first order behavior; instead, the observables behave as if they are in a continuous phase transition. In the same spirit as Sec.~\ref{sec:quantum}, we use the crossing point between $L=12$ and $L=24$ as an estimate for the phase boundary of the nematic phase. We think a specially tailored update scheme is necessary to ascertain the nature of phase transitions in the compass limit.

\section{Discussion \label{sec:discussion}}

To summarize, we have constructed the phase diagram for the bosonic Kitaev model by using the quantum Monte Carlo method. In particular, we show that the transitions from the paramagnetic phase to the various magnetic orders hosted in this system are generically fluctuation-induced first order transitions. This weakly first order transition is in contrast to the continuous phase transition in the square lattice $d^4$ magnet~\cite{Khaliullin2013,Jain2017,Souliou2017}, attesting to the rich physics displayed by $4d^4/5d^4$ magnets .

As we have shown in Sec.~\ref{sec:Landau}, the fluctuation-induced first order transition in the bosonic Kitaev model is a direct consequence of the $Z^{\times 3}_2$ internal symmetry and the lattice symmetry, which, together, strongly constrains the Landau theory. However, while the lattice symmetry is a robust feature of the honeycomb lattice, the $Z^{\times 3}_2$ internal symmetry is unique to the specific form of the Hamiltonian Eq.~\eqref{eq:Hamiltonian}. Many perturbations, allowed by the lattice symmetry, can break the $Z^{\times 3}_2$ internal symmetry: For instance, the higher order terms in the exchange interaction involving three triplon operators remove the internal symmetry all together~\cite{Khaliullin2013}. Meanwhile, trigonal distortions of the oxygen octahedra can induce a hybridization between triplons of different flavors, thereby reducing the $Z_2^{\times 3}$ symmetry to at most $Z_2$. 

It is therefore natural to enquire the fate of the fluctuation-induced first order transitions under these symmetry lowering perturbations. On general grounds, we expect that the first order transitions are robust against sufficiently weak perturbations, but may give way to transitions of other nature when the perturbations exceed a certain threshold. Extensive quantum Monte Carlo simulations are necessary to answer this question quantitatively. On this front, the trigonal distortion may represent a tractable case in that the perturbed minimal model Hamiltonian remains free from sign problems in the computation basis used in this work. 

Another important perturbation to the Hamiltonian is the external magnetic field. The system couples to the external field through the van Vleck magnetic moments as well as the triplon magnetic moments, both of which can drive the condensation of triplons. As the field breaks all the symmetries of the Hamiltonian, we expect that the transition will become a crossover unless the field is aligned in high symmetry directions. For instance, if the field is aligned in the $[111]$ direction, the magnetic phase transition may become the three-state Potts universality. 


Finally, we think it might be interesting to explore the dynamics of triplons in the bosonic Kitaev model. An experimental signature for the exchange-induced singlet magnetism is the soft amplitude modes in the dynamical spin structure factor in the magnetically ordered state. Such modes have been observed in the square lattice $d^4$ magnets~\cite{Jain2017,Souliou2017}. We expect that the frustrated dynamics of the triplons, unique to the honeycomb $d^4$ magnet, may produce unique features in the dynamic spin structure factor.

\begin{acknowledgments}
We thank Ji\v{r}\'{i} Chaloupka and Zhengxin Liu for helpful discussions. This work is supported by the Ministry of Science and Technology of China through the National Key R\&D Program of China (Grant No.~2022YFA1403800), the National Natural Science Foundation of China (Grant No.~11974396), and the Chinese Academy of Sciences through the Strategic Priority Research Program (Grant No.~XDB33020300) and the CAS Project for Young Scientists in Basic Research (Grant No.~YSBR-059). 
\end{acknowledgments}

\bibliography{bosonic_kitaev.bib}

\begin{thebibliography}{36}%
\makeatletter
\providecommand \@ifxundefined [1]{%
 \@ifx{#1\undefined}
}%
\providecommand \@ifnum [1]{%
 \ifnum #1\expandafter \@firstoftwo
 \else \expandafter \@secondoftwo
 \fi
}%
\providecommand \@ifx [1]{%
 \ifx #1\expandafter \@firstoftwo
 \else \expandafter \@secondoftwo
 \fi
}%
\providecommand \natexlab [1]{#1}%
\providecommand \enquote  [1]{``#1''}%
\providecommand \bibnamefont  [1]{#1}%
\providecommand \bibfnamefont [1]{#1}%
\providecommand \citenamefont [1]{#1}%
\providecommand \href@noop [0]{\@secondoftwo}%
\providecommand \href [0]{\begingroup \@sanitize@url \@href}%
\providecommand \@href[1]{\@@startlink{#1}\@@href}%
\providecommand \@@href[1]{\endgroup#1\@@endlink}%
\providecommand \@sanitize@url [0]{\catcode `\\12\catcode `\$12\catcode
  `\&12\catcode `\#12\catcode `\^12\catcode `\_12\catcode `\%12\relax}%
\providecommand \@@startlink[1]{}%
\providecommand \@@endlink[0]{}%
\providecommand \url  [0]{\begingroup\@sanitize@url \@url }%
\providecommand \@url [1]{\endgroup\@href {#1}{\urlprefix }}%
\providecommand \urlprefix  [0]{URL }%
\providecommand \Eprint [0]{\href }%
\providecommand \doibase [0]{https://doi.org/}%
\providecommand \selectlanguage [0]{\@gobble}%
\providecommand \bibinfo  [0]{\@secondoftwo}%
\providecommand \bibfield  [0]{\@secondoftwo}%
\providecommand \translation [1]{[#1]}%
\providecommand \BibitemOpen [0]{}%
\providecommand \bibitemStop [0]{}%
\providecommand \bibitemNoStop [0]{.\EOS\space}%
\providecommand \EOS [0]{\spacefactor3000\relax}%
\providecommand \BibitemShut  [1]{\csname bibitem#1\endcsname}%
\let\auto@bib@innerbib\@empty
\bibitem [{\citenamefont {Jackeli}\ and\ \citenamefont
  {Khaliullin}(2009)}]{Jackeli2009}%
  \BibitemOpen
  \bibfield  {author} {\bibinfo {author} {\bibfnamefont {G.}~\bibnamefont
  {Jackeli}}\ and\ \bibinfo {author} {\bibfnamefont {G.}~\bibnamefont
  {Khaliullin}},\ }\bibfield  {title} {\bibinfo {title} {{Mott Insulators in
  the Strong Spin-Orbit Coupling Limit: From Heisenberg to a Quantum Compass
  and Kitaev Models}},\ }\href {https://doi.org/10.1103/PhysRevLett.102.017205}
  {\bibfield  {journal} {\bibinfo  {journal} {Phys. Rev. Lett.}\ }\textbf
  {\bibinfo {volume} {102}},\ \bibinfo {pages} {017205} (\bibinfo {year}
  {2009})}\BibitemShut {NoStop}%
\bibitem [{\citenamefont {Chaloupka}\ \emph {et~al.}(2010)\citenamefont
  {Chaloupka}, \citenamefont {Jackeli},\ and\ \citenamefont
  {Khaliullin}}]{Chaloupka2010}%
  \BibitemOpen
  \bibfield  {author} {\bibinfo {author} {\bibfnamefont {J.}~\bibnamefont
  {Chaloupka}}, \bibinfo {author} {\bibfnamefont {G.}~\bibnamefont {Jackeli}},\
  and\ \bibinfo {author} {\bibfnamefont {G.}~\bibnamefont {Khaliullin}},\
  }\bibfield  {title} {\bibinfo {title} {{Kitaev-Heisenberg Model on a
  Honeycomb Lattice: Possible Exotic Phases in Iridium Oxides
  ${A}_{2}{\mathrm{IrO}}_{3}$}},\ }\href
  {https://doi.org/10.1103/PhysRevLett.105.027204} {\bibfield  {journal}
  {\bibinfo  {journal} {Phys. Rev. Lett.}\ }\textbf {\bibinfo {volume} {105}},\
  \bibinfo {pages} {027204} (\bibinfo {year} {2010})}\BibitemShut {NoStop}%
\bibitem [{\citenamefont {Chen}\ \emph {et~al.}(2010)\citenamefont {Chen},
  \citenamefont {Pereira},\ and\ \citenamefont {Balents}}]{Chen2010}%
  \BibitemOpen
  \bibfield  {author} {\bibinfo {author} {\bibfnamefont {G.}~\bibnamefont
  {Chen}}, \bibinfo {author} {\bibfnamefont {R.}~\bibnamefont {Pereira}},\ and\
  \bibinfo {author} {\bibfnamefont {L.}~\bibnamefont {Balents}},\ }\bibfield
  {title} {\bibinfo {title} {{Exotic phases induced by strong spin-orbit
  coupling in ordered double perovskites}},\ }\href
  {https://doi.org/10.1103/PhysRevB.82.174440} {\bibfield  {journal} {\bibinfo
  {journal} {Phys. Rev. B}\ }\textbf {\bibinfo {volume} {82}},\ \bibinfo
  {pages} {174440} (\bibinfo {year} {2010})}\BibitemShut {NoStop}%
\bibitem [{\citenamefont {Khaliullin}(2013)}]{Khaliullin2013}%
  \BibitemOpen
  \bibfield  {author} {\bibinfo {author} {\bibfnamefont {G.}~\bibnamefont
  {Khaliullin}},\ }\bibfield  {title} {\bibinfo {title} {{Excitonic Magnetism
  in Van Vleck--type ${d}^{4}$ Mott Insulators}},\ }\href
  {https://doi.org/10.1103/PhysRevLett.111.197201} {\bibfield  {journal}
  {\bibinfo  {journal} {Phys. Rev. Lett.}\ }\textbf {\bibinfo {volume} {111}},\
  \bibinfo {pages} {197201} (\bibinfo {year} {2013})}\BibitemShut {NoStop}%
\bibitem [{\citenamefont {Witczak-Krempa}\ \emph {et~al.}(2014)\citenamefont
  {Witczak-Krempa}, \citenamefont {Chen}, \citenamefont {Kim},\ and\
  \citenamefont {Balents}}]{Witczak2014}%
  \BibitemOpen
  \bibfield  {author} {\bibinfo {author} {\bibfnamefont {W.}~\bibnamefont
  {Witczak-Krempa}}, \bibinfo {author} {\bibfnamefont {G.}~\bibnamefont
  {Chen}}, \bibinfo {author} {\bibfnamefont {Y.~B.}\ \bibnamefont {Kim}},\ and\
  \bibinfo {author} {\bibfnamefont {L.}~\bibnamefont {Balents}},\ }\bibfield
  {title} {\bibinfo {title} {{Correlated Quantum Phenomena in the Strong
  Spin-Orbit Regime}},\ }\href
  {https://doi.org/10.1146/annurev-conmatphys-020911-125138} {\bibfield
  {journal} {\bibinfo  {journal} {Annual Review of Condensed Matter Physics}\
  }\textbf {\bibinfo {volume} {5}},\ \bibinfo {pages} {57} (\bibinfo {year}
  {2014})}\BibitemShut {NoStop}%
\bibitem [{\citenamefont {Meetei}\ \emph {et~al.}(2015)\citenamefont {Meetei},
  \citenamefont {Cole}, \citenamefont {Randeria},\ and\ \citenamefont
  {Trivedi}}]{Meetei2015}%
  \BibitemOpen
  \bibfield  {author} {\bibinfo {author} {\bibfnamefont {O.~N.}\ \bibnamefont
  {Meetei}}, \bibinfo {author} {\bibfnamefont {W.~S.}\ \bibnamefont {Cole}},
  \bibinfo {author} {\bibfnamefont {M.}~\bibnamefont {Randeria}},\ and\
  \bibinfo {author} {\bibfnamefont {N.}~\bibnamefont {Trivedi}},\ }\bibfield
  {title} {\bibinfo {title} {{Novel magnetic state in ${d}^{4}$ Mott
  insulators}},\ }\href {https://doi.org/10.1103/PhysRevB.91.054412} {\bibfield
   {journal} {\bibinfo  {journal} {Phys. Rev. B}\ }\textbf {\bibinfo {volume}
  {91}},\ \bibinfo {pages} {054412} (\bibinfo {year} {2015})}\BibitemShut
  {NoStop}%
\bibitem [{\citenamefont {Rau}\ \emph {et~al.}(2016)\citenamefont {Rau},
  \citenamefont {Lee},\ and\ \citenamefont {Kee}}]{Rau2016}%
  \BibitemOpen
  \bibfield  {author} {\bibinfo {author} {\bibfnamefont {J.~G.}\ \bibnamefont
  {Rau}}, \bibinfo {author} {\bibfnamefont {E.~K.-H.}\ \bibnamefont {Lee}},\
  and\ \bibinfo {author} {\bibfnamefont {H.-Y.}\ \bibnamefont {Kee}},\
  }\bibfield  {title} {\bibinfo {title} {{Spin-Orbit Physics Giving Rise to
  Novel Phases in Correlated Systems: Iridates and Related Materials}},\ }\href
  {https://doi.org/10.1146/annurev-conmatphys-031115-011319} {\bibfield
  {journal} {\bibinfo  {journal} {Annual Review of Condensed Matter Physics}\
  }\textbf {\bibinfo {volume} {7}},\ \bibinfo {pages} {195} (\bibinfo {year}
  {2016})}\BibitemShut {NoStop}%
\bibitem [{\citenamefont {Winter}\ \emph {et~al.}(2017)\citenamefont {Winter},
  \citenamefont {Tsirlin}, \citenamefont {Daghofer}, \citenamefont {van~den
  Brink}, \citenamefont {Singh}, \citenamefont {Gegenwart},\ and\ \citenamefont
  {Valent{\'\i}}}]{Winter2017}%
  \BibitemOpen
  \bibfield  {author} {\bibinfo {author} {\bibfnamefont {S.~M.}\ \bibnamefont
  {Winter}}, \bibinfo {author} {\bibfnamefont {A.~A.}\ \bibnamefont {Tsirlin}},
  \bibinfo {author} {\bibfnamefont {M.}~\bibnamefont {Daghofer}}, \bibinfo
  {author} {\bibfnamefont {J.}~\bibnamefont {van~den Brink}}, \bibinfo {author}
  {\bibfnamefont {Y.}~\bibnamefont {Singh}}, \bibinfo {author} {\bibfnamefont
  {P.}~\bibnamefont {Gegenwart}},\ and\ \bibinfo {author} {\bibfnamefont
  {R.}~\bibnamefont {Valent{\'\i}}},\ }\bibfield  {title} {\bibinfo {title}
  {{Models and materials for generalized Kitaev magnetism}},\ }\href
  {https://doi.org/10.1088/1361-648X/aa8cf5} {\bibfield  {journal} {\bibinfo
  {journal} {Journal of Physics: Condensed Matter}\ }\textbf {\bibinfo {volume}
  {29}},\ \bibinfo {pages} {493002} (\bibinfo {year} {2017})}\BibitemShut
  {NoStop}%
\bibitem [{\citenamefont {Takayama}\ \emph {et~al.}(2021)\citenamefont
  {Takayama}, \citenamefont {Chaloupka}, \citenamefont {Smerald}, \citenamefont
  {Khaliullin},\ and\ \citenamefont {Takagi}}]{Takayama2021}%
  \BibitemOpen
  \bibfield  {author} {\bibinfo {author} {\bibfnamefont {T.}~\bibnamefont
  {Takayama}}, \bibinfo {author} {\bibfnamefont {J.}~\bibnamefont {Chaloupka}},
  \bibinfo {author} {\bibfnamefont {A.}~\bibnamefont {Smerald}}, \bibinfo
  {author} {\bibfnamefont {G.}~\bibnamefont {Khaliullin}},\ and\ \bibinfo
  {author} {\bibfnamefont {H.}~\bibnamefont {Takagi}},\ }\bibfield  {title}
  {\bibinfo {title} {{Spin--Orbit-Entangled Electronic Phases in 4d and 5d
  Transition-Metal Compounds}},\ }\href
  {https://doi.org/10.7566/JPSJ.90.062001} {\bibfield  {journal} {\bibinfo
  {journal} {Journal of the Physical Society of Japan}\ }\textbf {\bibinfo
  {volume} {90}},\ \bibinfo {pages} {062001} (\bibinfo {year}
  {2021})}\BibitemShut {NoStop}%
\bibitem [{\citenamefont {Trebst}\ and\ \citenamefont
  {Hickey}(2022)}]{Trebst2022}%
  \BibitemOpen
  \bibfield  {author} {\bibinfo {author} {\bibfnamefont {S.}~\bibnamefont
  {Trebst}}\ and\ \bibinfo {author} {\bibfnamefont {C.}~\bibnamefont
  {Hickey}},\ }\bibfield  {title} {\bibinfo {title} {Kitaev materials},\ }\href
  {https://doi.org/https://doi.org/10.1016/j.physrep.2021.11.003} {\bibfield
  {journal} {\bibinfo  {journal} {Physics Reports}\ }\textbf {\bibinfo {volume}
  {950}},\ \bibinfo {pages} {1} (\bibinfo {year} {2022})},\ \bibinfo {note}
  {kitaev materials}\BibitemShut {NoStop}%
\bibitem [{\citenamefont {Anisimov}\ \emph {et~al.}(2019)\citenamefont
  {Anisimov}, \citenamefont {Aust}, \citenamefont {Khaliullin},\ and\
  \citenamefont {Daghofer}}]{Anisimov2019}%
  \BibitemOpen
  \bibfield  {author} {\bibinfo {author} {\bibfnamefont {P.~S.}\ \bibnamefont
  {Anisimov}}, \bibinfo {author} {\bibfnamefont {F.}~\bibnamefont {Aust}},
  \bibinfo {author} {\bibfnamefont {G.}~\bibnamefont {Khaliullin}},\ and\
  \bibinfo {author} {\bibfnamefont {M.}~\bibnamefont {Daghofer}},\ }\bibfield
  {title} {\bibinfo {title} {{Nontrivial Triplon Topology and Triplon Liquid in
  Kitaev-Heisenberg-type Excitonic Magnets}},\ }\href
  {https://doi.org/10.1103/PhysRevLett.122.177201} {\bibfield  {journal}
  {\bibinfo  {journal} {Phys. Rev. Lett.}\ }\textbf {\bibinfo {volume} {122}},\
  \bibinfo {pages} {177201} (\bibinfo {year} {2019})}\BibitemShut {NoStop}%
\bibitem [{\citenamefont {Chaloupka}\ and\ \citenamefont
  {Khaliullin}(2019)}]{Chaloupka2019}%
  \BibitemOpen
  \bibfield  {author} {\bibinfo {author} {\bibfnamefont {J.}~\bibnamefont
  {Chaloupka}}\ and\ \bibinfo {author} {\bibfnamefont {G.}~\bibnamefont
  {Khaliullin}},\ }\bibfield  {title} {\bibinfo {title} {{Highly frustrated
  magnetism in relativistic ${d}^{4}$ Mott insulators: Bosonic analog of the
  Kitaev honeycomb model}},\ }\href
  {https://doi.org/10.1103/PhysRevB.100.224413} {\bibfield  {journal} {\bibinfo
   {journal} {Phys. Rev. B}\ }\textbf {\bibinfo {volume} {100}},\ \bibinfo
  {pages} {224413} (\bibinfo {year} {2019})}\BibitemShut {NoStop}%
\bibitem [{\citenamefont {Kimber}\ \emph {et~al.}(2010)\citenamefont {Kimber},
  \citenamefont {Ling}, \citenamefont {Morris}, \citenamefont {Chemseddine},
  \citenamefont {Henry},\ and\ \citenamefont {Argyriou}}]{Kimber2010}%
  \BibitemOpen
  \bibfield  {author} {\bibinfo {author} {\bibfnamefont {S.~A.~J.}\
  \bibnamefont {Kimber}}, \bibinfo {author} {\bibfnamefont {C.~D.}\
  \bibnamefont {Ling}}, \bibinfo {author} {\bibfnamefont {D.~J.~P.}\
  \bibnamefont {Morris}}, \bibinfo {author} {\bibfnamefont {A.}~\bibnamefont
  {Chemseddine}}, \bibinfo {author} {\bibfnamefont {P.~F.}\ \bibnamefont
  {Henry}},\ and\ \bibinfo {author} {\bibfnamefont {D.~N.}\ \bibnamefont
  {Argyriou}},\ }\bibfield  {title} {\bibinfo {title} {{Interlayer tuning of
  electronic and magnetic properties in honeycomb ordered
  Ag$_3$LiRu$_2$O$_6$}},\ }\href {https://doi.org/10.1039/C0JM00678E}
  {\bibfield  {journal} {\bibinfo  {journal} {J. Mater. Chem.}\ }\textbf
  {\bibinfo {volume} {20}},\ \bibinfo {pages} {8021} (\bibinfo {year}
  {2010})}\BibitemShut {NoStop}%
\bibitem [{\citenamefont {Kumar}\ \emph {et~al.}(2019)\citenamefont {Kumar},
  \citenamefont {Dey}, \citenamefont {Ette}, \citenamefont {Ramesha},
  \citenamefont {Chakraborty}, \citenamefont {Dasgupta}, \citenamefont {Orain},
  \citenamefont {Baines}, \citenamefont {T\'oth}, \citenamefont {Shahee},
  \citenamefont {Kundu}, \citenamefont {Prinz-Zwick}, \citenamefont {Gippius},
  \citenamefont {B\"uttgen}, \citenamefont {Gegenwart},\ and\ \citenamefont
  {Mahajan}}]{Kumar2019}%
  \BibitemOpen
  \bibfield  {author} {\bibinfo {author} {\bibfnamefont {R.}~\bibnamefont
  {Kumar}}, \bibinfo {author} {\bibfnamefont {T.}~\bibnamefont {Dey}}, \bibinfo
  {author} {\bibfnamefont {P.~M.}\ \bibnamefont {Ette}}, \bibinfo {author}
  {\bibfnamefont {K.}~\bibnamefont {Ramesha}}, \bibinfo {author} {\bibfnamefont
  {A.}~\bibnamefont {Chakraborty}}, \bibinfo {author} {\bibfnamefont
  {I.}~\bibnamefont {Dasgupta}}, \bibinfo {author} {\bibfnamefont {J.~C.}\
  \bibnamefont {Orain}}, \bibinfo {author} {\bibfnamefont {C.}~\bibnamefont
  {Baines}}, \bibinfo {author} {\bibfnamefont {S.}~\bibnamefont {T\'oth}},
  \bibinfo {author} {\bibfnamefont {A.}~\bibnamefont {Shahee}}, \bibinfo
  {author} {\bibfnamefont {S.}~\bibnamefont {Kundu}}, \bibinfo {author}
  {\bibfnamefont {M.}~\bibnamefont {Prinz-Zwick}}, \bibinfo {author}
  {\bibfnamefont {A.~A.}\ \bibnamefont {Gippius}}, \bibinfo {author}
  {\bibfnamefont {N.}~\bibnamefont {B\"uttgen}}, \bibinfo {author}
  {\bibfnamefont {P.}~\bibnamefont {Gegenwart}},\ and\ \bibinfo {author}
  {\bibfnamefont {A.~V.}\ \bibnamefont {Mahajan}},\ }\bibfield  {title}
  {\bibinfo {title} {{Unconventional magnetism in the $4{d}^{4}$-based $S=1$
  honeycomb system ${\mathrm{Ag}}_{3}{\mathrm{LiRu}}_{2}{\mathrm{O}}_{6}$}},\
  }\href {https://doi.org/10.1103/PhysRevB.99.054417} {\bibfield  {journal}
  {\bibinfo  {journal} {Phys. Rev. B}\ }\textbf {\bibinfo {volume} {99}},\
  \bibinfo {pages} {054417} (\bibinfo {year} {2019})}\BibitemShut {NoStop}%
\bibitem [{\citenamefont {Mogare}\ \emph {et~al.}(2004)\citenamefont {Mogare},
  \citenamefont {Friese}, \citenamefont {Klein},\ and\ \citenamefont
  {Jansen}}]{Mogare2004}%
  \BibitemOpen
  \bibfield  {author} {\bibinfo {author} {\bibfnamefont {K.~M.}\ \bibnamefont
  {Mogare}}, \bibinfo {author} {\bibfnamefont {K.}~\bibnamefont {Friese}},
  \bibinfo {author} {\bibfnamefont {W.}~\bibnamefont {Klein}},\ and\ \bibinfo
  {author} {\bibfnamefont {M.}~\bibnamefont {Jansen}},\ }\bibfield  {title}
  {\bibinfo {title} {{Syntheses and Crystal Structures of Two Sodium
  Ruthenates: Na$_2$RuO$_4$ and Na$_2$RuO$_3$}},\ }\href
  {https://doi.org/https://doi.org/10.1002/zaac.200400012} {\bibfield
  {journal} {\bibinfo  {journal} {Zeitschrift f{\"u}r anorganische und
  allgemeine Chemie}\ }\textbf {\bibinfo {volume} {630}},\ \bibinfo {pages}
  {547} (\bibinfo {year} {2004})}\BibitemShut {NoStop}%
\bibitem [{\citenamefont {Wang}\ \emph {et~al.}(2014)\citenamefont {Wang},
  \citenamefont {Terzic}, \citenamefont {Qi}, \citenamefont {Ye}, \citenamefont
  {Yuan}, \citenamefont {Aswartham}, \citenamefont {Streltsov}, \citenamefont
  {Khomskii}, \citenamefont {Kaul},\ and\ \citenamefont {Cao}}]{Wang2014}%
  \BibitemOpen
  \bibfield  {author} {\bibinfo {author} {\bibfnamefont {J.~C.}\ \bibnamefont
  {Wang}}, \bibinfo {author} {\bibfnamefont {J.}~\bibnamefont {Terzic}},
  \bibinfo {author} {\bibfnamefont {T.~F.}\ \bibnamefont {Qi}}, \bibinfo
  {author} {\bibfnamefont {F.}~\bibnamefont {Ye}}, \bibinfo {author}
  {\bibfnamefont {S.~J.}\ \bibnamefont {Yuan}}, \bibinfo {author}
  {\bibfnamefont {S.}~\bibnamefont {Aswartham}}, \bibinfo {author}
  {\bibfnamefont {S.~V.}\ \bibnamefont {Streltsov}}, \bibinfo {author}
  {\bibfnamefont {D.~I.}\ \bibnamefont {Khomskii}}, \bibinfo {author}
  {\bibfnamefont {R.~K.}\ \bibnamefont {Kaul}},\ and\ \bibinfo {author}
  {\bibfnamefont {G.}~\bibnamefont {Cao}},\ }\bibfield  {title} {\bibinfo
  {title} {{Lattice-tuned magnetism of ${\mathrm{Ru}}^{4+}(4{d}^{4})$ ions in
  single crystals of the layered honeycomb ruthenates
  ${\mathrm{Li}}_{2}{\mathrm{RuO}}_{3}$ and
  ${\mathrm{Na}}_{2}{\mathrm{RuO}}_{3}$}},\ }\href
  {https://doi.org/10.1103/PhysRevB.90.161110} {\bibfield  {journal} {\bibinfo
  {journal} {Phys. Rev. B}\ }\textbf {\bibinfo {volume} {90}},\ \bibinfo
  {pages} {161110} (\bibinfo {year} {2014})}\BibitemShut {NoStop}%
\bibitem [{\citenamefont {Gapontsev}\ \emph {et~al.}(2017)\citenamefont
  {Gapontsev}, \citenamefont {Kurmaev}, \citenamefont {Sathish}, \citenamefont
  {Yun}, \citenamefont {Park},\ and\ \citenamefont
  {Streltsov}}]{Gapontsev2017}%
  \BibitemOpen
  \bibfield  {author} {\bibinfo {author} {\bibfnamefont {V.~V.}\ \bibnamefont
  {Gapontsev}}, \bibinfo {author} {\bibfnamefont {E.~Z.}\ \bibnamefont
  {Kurmaev}}, \bibinfo {author} {\bibfnamefont {C.~I.}\ \bibnamefont
  {Sathish}}, \bibinfo {author} {\bibfnamefont {S.}~\bibnamefont {Yun}},
  \bibinfo {author} {\bibfnamefont {J.-G.}\ \bibnamefont {Park}},\ and\
  \bibinfo {author} {\bibfnamefont {S.~V.}\ \bibnamefont {Streltsov}},\
  }\bibfield  {title} {\bibinfo {title} {{Spectral and magnetic properties of
  Na$_2$RuO$_3$}},\ }\href {https://doi.org/10.1088/1361-648X/aa7fd6}
  {\bibfield  {journal} {\bibinfo  {journal} {Journal of Physics: Condensed
  Matter}\ }\textbf {\bibinfo {volume} {29}},\ \bibinfo {pages} {405804}
  (\bibinfo {year} {2017})}\BibitemShut {NoStop}%
\bibitem [{\citenamefont {Veiga}\ \emph {et~al.}(2020)\citenamefont {Veiga},
  \citenamefont {Etter}, \citenamefont {Cappelli}, \citenamefont {Jacobsen},
  \citenamefont {Vale}, \citenamefont {Dashwood}, \citenamefont {Le},
  \citenamefont {Baumberger}, \citenamefont {McMorrow},\ and\ \citenamefont
  {Perry}}]{Veiga2020}%
  \BibitemOpen
  \bibfield  {author} {\bibinfo {author} {\bibfnamefont {L.~S.~I.}\
  \bibnamefont {Veiga}}, \bibinfo {author} {\bibfnamefont {M.}~\bibnamefont
  {Etter}}, \bibinfo {author} {\bibfnamefont {E.}~\bibnamefont {Cappelli}},
  \bibinfo {author} {\bibfnamefont {H.}~\bibnamefont {Jacobsen}}, \bibinfo
  {author} {\bibfnamefont {J.~G.}\ \bibnamefont {Vale}}, \bibinfo {author}
  {\bibfnamefont {C.~D.}\ \bibnamefont {Dashwood}}, \bibinfo {author}
  {\bibfnamefont {D.}~\bibnamefont {Le}}, \bibinfo {author} {\bibfnamefont
  {F.}~\bibnamefont {Baumberger}}, \bibinfo {author} {\bibfnamefont {D.~F.}\
  \bibnamefont {McMorrow}},\ and\ \bibinfo {author} {\bibfnamefont {R.~S.}\
  \bibnamefont {Perry}},\ }\bibfield  {title} {\bibinfo {title} {{Correlated
  electron metal properties of the honeycomb ruthenate
  ${\mathrm{Na}}_{2}{\mathrm{RuO}}_{3}$}},\ }\href
  {https://doi.org/10.1103/PhysRevMaterials.4.094202} {\bibfield  {journal}
  {\bibinfo  {journal} {Phys. Rev. Materials}\ }\textbf {\bibinfo {volume}
  {4}},\ \bibinfo {pages} {094202} (\bibinfo {year} {2020})}\BibitemShut
  {NoStop}%
\bibitem [{\citenamefont {Song}\ \emph {et~al.}(2021)\citenamefont {Song},
  \citenamefont {Zhu}, \citenamefont {Yang}, \citenamefont {Wei}, \citenamefont
  {Zhang}, \citenamefont {Yang}, \citenamefont {Sheng}, \citenamefont {Qi},
  \citenamefont {Ni}, \citenamefont {Li}, \citenamefont {Li}, \citenamefont
  {Cao}, \citenamefont {Meng}, \citenamefont {Li}, \citenamefont {Shi},\ and\
  \citenamefont {Li}}]{Song2021}%
  \BibitemOpen
  \bibfield  {author} {\bibinfo {author} {\bibfnamefont {P.}~\bibnamefont
  {Song}}, \bibinfo {author} {\bibfnamefont {K.}~\bibnamefont {Zhu}}, \bibinfo
  {author} {\bibfnamefont {F.}~\bibnamefont {Yang}}, \bibinfo {author}
  {\bibfnamefont {Y.}~\bibnamefont {Wei}}, \bibinfo {author} {\bibfnamefont
  {L.}~\bibnamefont {Zhang}}, \bibinfo {author} {\bibfnamefont
  {H.}~\bibnamefont {Yang}}, \bibinfo {author} {\bibfnamefont {X.-L.}\
  \bibnamefont {Sheng}}, \bibinfo {author} {\bibfnamefont {Y.}~\bibnamefont
  {Qi}}, \bibinfo {author} {\bibfnamefont {J.}~\bibnamefont {Ni}}, \bibinfo
  {author} {\bibfnamefont {S.}~\bibnamefont {Li}}, \bibinfo {author}
  {\bibfnamefont {Y.}~\bibnamefont {Li}}, \bibinfo {author} {\bibfnamefont
  {G.}~\bibnamefont {Cao}}, \bibinfo {author} {\bibfnamefont {Z.~Y.}\
  \bibnamefont {Meng}}, \bibinfo {author} {\bibfnamefont {W.}~\bibnamefont
  {Li}}, \bibinfo {author} {\bibfnamefont {Y.}~\bibnamefont {Shi}},\ and\
  \bibinfo {author} {\bibfnamefont {S.}~\bibnamefont {Li}},\ }\bibfield
  {title} {\bibinfo {title} {{Evidence for the random singlet phase in the
  honeycomb iridate ${\mathrm{SrIr}}_{2}{\mathrm{O}}_{6}$}},\ }\href
  {https://doi.org/10.1103/PhysRevB.103.L241114} {\bibfield  {journal}
  {\bibinfo  {journal} {Phys. Rev. B}\ }\textbf {\bibinfo {volume} {103}},\
  \bibinfo {pages} {L241114} (\bibinfo {year} {2021})}\BibitemShut {NoStop}%
\bibitem [{\citenamefont {Giamarchi}\ \emph {et~al.}(2008)\citenamefont
  {Giamarchi}, \citenamefont {R{\"u}egg},\ and\ \citenamefont
  {Tchernyshyov}}]{Giamarchi2008}%
  \BibitemOpen
  \bibfield  {author} {\bibinfo {author} {\bibfnamefont {T.}~\bibnamefont
  {Giamarchi}}, \bibinfo {author} {\bibfnamefont {C.}~\bibnamefont
  {R{\"u}egg}},\ and\ \bibinfo {author} {\bibfnamefont {O.}~\bibnamefont
  {Tchernyshyov}},\ }\bibfield  {title} {\bibinfo {title} {{Bose--Einstein
  condensation in magnetic insulators}},\ }\href
  {https://doi.org/10.1038/nphys893} {\bibfield  {journal} {\bibinfo  {journal}
  {Nature Physics}\ }\textbf {\bibinfo {volume} {4}},\ \bibinfo {pages} {198}
  (\bibinfo {year} {2008})}\BibitemShut {NoStop}%
\bibitem [{\citenamefont {Cooper}(1972)}]{Cooper1972}%
  \BibitemOpen
  \bibfield  {author} {\bibinfo {author} {\bibfnamefont {B.~R.}\ \bibnamefont
  {Cooper}},\ }\bibinfo {title} {{Phenomenological Theory of Magnetic Ordering:
  Importance of Interactions with the Crystal Lattice}},\ in\ \href
  {https://doi.org/10.1007/978-1-4757-5691-3_2} {\emph {\bibinfo {booktitle}
  {{Magnetic Properties of Rare Earth Metals}}}},\ \bibinfo {editor} {edited
  by\ \bibinfo {editor} {\bibfnamefont {R.~J.}\ \bibnamefont {Elliott}}}\
  (\bibinfo  {publisher} {Springer US},\ \bibinfo {address} {Boston, MA},\
  \bibinfo {year} {1972})\ pp.\ \bibinfo {pages} {17--80}\BibitemShut {NoStop}%
\bibitem [{\citenamefont {Kitaev}(2006)}]{Kitaev2006}%
  \BibitemOpen
  \bibfield  {author} {\bibinfo {author} {\bibfnamefont {A.}~\bibnamefont
  {Kitaev}},\ }\bibfield  {title} {\bibinfo {title} {{Anyons in an exactly
  solved model and beyond}},\ }\href
  {https://doi.org/https://doi.org/10.1016/j.aop.2005.10.005} {\bibfield
  {journal} {\bibinfo  {journal} {Annals of Physics}\ }\textbf {\bibinfo
  {volume} {321}},\ \bibinfo {pages} {2} (\bibinfo {year} {2006})},\ \bibinfo
  {note} {january Special Issue}\BibitemShut {NoStop}%
\bibitem [{\citenamefont {Aharony}(1976)}]{Aharony1976}%
  \BibitemOpen
  \bibfield  {author} {\bibinfo {author} {\bibfnamefont {A.}~\bibnamefont
  {Aharony}},\ }\bibfield  {title} {\bibinfo {title} {{Dependence of universal
  critical behavior on symmetry and range of interaction}},\ }in\ \href@noop {}
  {\emph {\bibinfo {booktitle} {Phase Transitions and Critical Phenomena}}},\
  Vol.~\bibinfo {volume} {6},\ \bibinfo {editor} {edited by\ \bibinfo {editor}
  {\bibfnamefont {C.}~\bibnamefont {Domb}}\ and\ \bibinfo {editor}
  {\bibfnamefont {M.~S.}\ \bibnamefont {Green}}}\ (\bibinfo  {publisher}
  {Academic Press},\ \bibinfo {address} {New York},\ \bibinfo {year} {1976})\
  pp.\ \bibinfo {pages} {357--424}\BibitemShut {NoStop}%
\bibitem [{\citenamefont {Nienhuis}\ \emph {et~al.}(1983)\citenamefont
  {Nienhuis}, \citenamefont {Riedel},\ and\ \citenamefont
  {Schick}}]{Nienhuis1983}%
  \BibitemOpen
  \bibfield  {author} {\bibinfo {author} {\bibfnamefont {B.}~\bibnamefont
  {Nienhuis}}, \bibinfo {author} {\bibfnamefont {E.~K.}\ \bibnamefont
  {Riedel}},\ and\ \bibinfo {author} {\bibfnamefont {M.}~\bibnamefont
  {Schick}},\ }\bibfield  {title} {\bibinfo {title} {{Critical behavior of the
  $n$-component cubic model and the Ashkin-Teller fixed line}},\ }\href
  {https://doi.org/10.1103/PhysRevB.27.5625} {\bibfield  {journal} {\bibinfo
  {journal} {Phys. Rev. B}\ }\textbf {\bibinfo {volume} {27}},\ \bibinfo
  {pages} {5625} (\bibinfo {year} {1983})}\BibitemShut {NoStop}%
\bibitem [{\citenamefont {Roelofs}\ and\ \citenamefont
  {Jackson}(1993)}]{Roelofs1993}%
  \BibitemOpen
  \bibfield  {author} {\bibinfo {author} {\bibfnamefont {L.~D.}\ \bibnamefont
  {Roelofs}}\ and\ \bibinfo {author} {\bibfnamefont {C.}~\bibnamefont
  {Jackson}},\ }\bibfield  {title} {\bibinfo {title} {{Critical and finite-size
  behavior of the Heisenberg model with face-centered-cubic anisotropy}},\
  }\href {https://doi.org/10.1103/PhysRevB.47.197} {\bibfield  {journal}
  {\bibinfo  {journal} {Phys. Rev. B}\ }\textbf {\bibinfo {volume} {47}},\
  \bibinfo {pages} {197} (\bibinfo {year} {1993})}\BibitemShut {NoStop}%
\bibitem [{\citenamefont {Cardy}(1996)}]{Cardy1996}%
  \BibitemOpen
  \bibfield  {author} {\bibinfo {author} {\bibfnamefont {J.}~\bibnamefont
  {Cardy}},\ }\bibinfo {title} {{Scaling and Renormalization in Statistical
  Physics}}\ (\bibinfo  {publisher} {Cambridge University Press},\ \bibinfo
  {year} {1996})\ Chap.\ \bibinfo {chapter} {5.8}\BibitemShut {NoStop}%
\bibitem [{\citenamefont {Dorier}\ \emph {et~al.}(2005)\citenamefont {Dorier},
  \citenamefont {Becca},\ and\ \citenamefont {Mila}}]{Dorier2005}%
  \BibitemOpen
  \bibfield  {author} {\bibinfo {author} {\bibfnamefont {J.}~\bibnamefont
  {Dorier}}, \bibinfo {author} {\bibfnamefont {F.}~\bibnamefont {Becca}},\ and\
  \bibinfo {author} {\bibfnamefont {F.}~\bibnamefont {Mila}},\ }\bibfield
  {title} {\bibinfo {title} {Quantum compass model on the square lattice},\
  }\href {https://doi.org/10.1103/PhysRevB.72.024448} {\bibfield  {journal}
  {\bibinfo  {journal} {Phys. Rev. B}\ }\textbf {\bibinfo {volume} {72}},\
  \bibinfo {pages} {024448} (\bibinfo {year} {2005})}\BibitemShut {NoStop}%
\bibitem [{\citenamefont {Wenzel}\ and\ \citenamefont
  {Janke}(2008)}]{Wenzel2008}%
  \BibitemOpen
  \bibfield  {author} {\bibinfo {author} {\bibfnamefont {S.}~\bibnamefont
  {Wenzel}}\ and\ \bibinfo {author} {\bibfnamefont {W.}~\bibnamefont {Janke}},\
  }\bibfield  {title} {\bibinfo {title} {{Monte Carlo simulations of the
  directional-ordering transition in the two-dimensional classical and quantum
  compass model}},\ }\href {https://doi.org/10.1103/PhysRevB.78.064402}
  {\bibfield  {journal} {\bibinfo  {journal} {Phys. Rev. B}\ }\textbf {\bibinfo
  {volume} {78}},\ \bibinfo {pages} {064402} (\bibinfo {year}
  {2008})}\BibitemShut {NoStop}%
\bibitem [{\citenamefont {Nussinov}\ and\ \citenamefont {van~den
  Brink}(2015)}]{Nussinov2015}%
  \BibitemOpen
  \bibfield  {author} {\bibinfo {author} {\bibfnamefont {Z.}~\bibnamefont
  {Nussinov}}\ and\ \bibinfo {author} {\bibfnamefont {J.}~\bibnamefont {van~den
  Brink}},\ }\bibfield  {title} {\bibinfo {title} {{Compass models: Theory and
  physical motivations}},\ }\href {https://doi.org/10.1103/RevModPhys.87.1}
  {\bibfield  {journal} {\bibinfo  {journal} {Rev. Mod. Phys.}\ }\textbf
  {\bibinfo {volume} {87}},\ \bibinfo {pages} {1} (\bibinfo {year}
  {2015})}\BibitemShut {NoStop}%
\bibitem [{\citenamefont {Chaloupka}\ and\ \citenamefont
  {Khaliullin}(2015)}]{Chaloupka2016}%
  \BibitemOpen
  \bibfield  {author} {\bibinfo {author} {\bibfnamefont {J.}~\bibnamefont
  {Chaloupka}}\ and\ \bibinfo {author} {\bibfnamefont {G.}~\bibnamefont
  {Khaliullin}},\ }\bibfield  {title} {\bibinfo {title} {{Hidden symmetries of
  the extended Kitaev-Heisenberg model: Implications for the honeycomb-lattice
  iridates ${A}_{2}{\mathrm{IrO}}_{3}$}},\ }\href
  {https://doi.org/10.1103/PhysRevB.92.024413} {\bibfield  {journal} {\bibinfo
  {journal} {Phys. Rev. B}\ }\textbf {\bibinfo {volume} {92}},\ \bibinfo
  {pages} {024413} (\bibinfo {year} {2015})}\BibitemShut {NoStop}%
\bibitem [{\citenamefont {Sylju\aa{}sen}\ and\ \citenamefont
  {Sandvik}(2002)}]{Syljuasen2002}%
  \BibitemOpen
  \bibfield  {author} {\bibinfo {author} {\bibfnamefont {O.~F.}\ \bibnamefont
  {Sylju\aa{}sen}}\ and\ \bibinfo {author} {\bibfnamefont {A.~W.}\ \bibnamefont
  {Sandvik}},\ }\bibfield  {title} {\bibinfo {title} {{Quantum Monte Carlo with
  directed loops}},\ }\href {https://doi.org/10.1103/PhysRevE.66.046701}
  {\bibfield  {journal} {\bibinfo  {journal} {Phys. Rev. E}\ }\textbf {\bibinfo
  {volume} {66}},\ \bibinfo {pages} {046701} (\bibinfo {year}
  {2002})}\BibitemShut {NoStop}%
\bibitem [{\citenamefont {Sandvik}(2010)}]{Sandvik2010}%
  \BibitemOpen
  \bibfield  {author} {\bibinfo {author} {\bibfnamefont {A.~W.}\ \bibnamefont
  {Sandvik}},\ }\bibfield  {title} {\bibinfo {title} {{Computational Studies of
  Quantum Spin Systems}},\ }\href {https://doi.org/10.1063/1.3518900}
  {\bibfield  {journal} {\bibinfo  {journal} {AIP Conference Proceedings}\
  }\textbf {\bibinfo {volume} {1297}},\ \bibinfo {pages} {135} (\bibinfo {year}
  {2010})}\BibitemShut {NoStop}%
\bibitem [{\citenamefont {Binder}\ and\ \citenamefont
  {Landau}(1984)}]{Binder1984}%
  \BibitemOpen
  \bibfield  {author} {\bibinfo {author} {\bibfnamefont {K.}~\bibnamefont
  {Binder}}\ and\ \bibinfo {author} {\bibfnamefont {D.~P.}\ \bibnamefont
  {Landau}},\ }\bibfield  {title} {\bibinfo {title} {Finite-size scaling at
  first-order phase transitions},\ }\href
  {https://doi.org/10.1103/PhysRevB.30.1477} {\bibfield  {journal} {\bibinfo
  {journal} {Phys. Rev. B}\ }\textbf {\bibinfo {volume} {30}},\ \bibinfo
  {pages} {1477} (\bibinfo {year} {1984})}\BibitemShut {NoStop}%
\bibitem [{\citenamefont {Lee}\ and\ \citenamefont
  {Kosterlitz}(1991)}]{Lee1991}%
  \BibitemOpen
  \bibfield  {author} {\bibinfo {author} {\bibfnamefont {J.}~\bibnamefont
  {Lee}}\ and\ \bibinfo {author} {\bibfnamefont {J.~M.}\ \bibnamefont
  {Kosterlitz}},\ }\bibfield  {title} {\bibinfo {title} {{Three-dimensional
  q-state Potts model: Monte Carlo study near q=3}},\ }\href
  {https://doi.org/10.1103/PhysRevB.43.1268} {\bibfield  {journal} {\bibinfo
  {journal} {Phys. Rev. B}\ }\textbf {\bibinfo {volume} {43}},\ \bibinfo
  {pages} {1268} (\bibinfo {year} {1991})}\BibitemShut {NoStop}%
\bibitem [{\citenamefont {Jain}\ \emph {et~al.}(2017)\citenamefont {Jain},
  \citenamefont {Krautloher}, \citenamefont {Porras}, \citenamefont {Ryu},
  \citenamefont {Chen}, \citenamefont {Abernathy}, \citenamefont {Park},
  \citenamefont {Ivanov}, \citenamefont {Chaloupka}, \citenamefont
  {Khaliullin}, \citenamefont {Keimer},\ and\ \citenamefont {Kim}}]{Jain2017}%
  \BibitemOpen
  \bibfield  {author} {\bibinfo {author} {\bibfnamefont {A.}~\bibnamefont
  {Jain}}, \bibinfo {author} {\bibfnamefont {M.}~\bibnamefont {Krautloher}},
  \bibinfo {author} {\bibfnamefont {J.}~\bibnamefont {Porras}}, \bibinfo
  {author} {\bibfnamefont {G.~H.}\ \bibnamefont {Ryu}}, \bibinfo {author}
  {\bibfnamefont {D.~P.}\ \bibnamefont {Chen}}, \bibinfo {author}
  {\bibfnamefont {D.~L.}\ \bibnamefont {Abernathy}}, \bibinfo {author}
  {\bibfnamefont {J.~T.}\ \bibnamefont {Park}}, \bibinfo {author}
  {\bibfnamefont {A.}~\bibnamefont {Ivanov}}, \bibinfo {author} {\bibfnamefont
  {J.}~\bibnamefont {Chaloupka}}, \bibinfo {author} {\bibfnamefont
  {G.}~\bibnamefont {Khaliullin}}, \bibinfo {author} {\bibfnamefont
  {B.}~\bibnamefont {Keimer}},\ and\ \bibinfo {author} {\bibfnamefont {B.~J.}\
  \bibnamefont {Kim}},\ }\bibfield  {title} {\bibinfo {title} {{Higgs mode and
  its decay in a two-dimensional antiferromagnet}},\ }\href
  {https://doi.org/10.1038/nphys4077} {\bibfield  {journal} {\bibinfo
  {journal} {Nature Physics}\ }\textbf {\bibinfo {volume} {13}},\ \bibinfo
  {pages} {633} (\bibinfo {year} {2017})}\BibitemShut {NoStop}%
\bibitem [{\citenamefont {Souliou}\ \emph {et~al.}(2017)\citenamefont
  {Souliou}, \citenamefont {Chaloupka}, \citenamefont {Khaliullin},
  \citenamefont {Ryu}, \citenamefont {Jain}, \citenamefont {Kim}, \citenamefont
  {Le~Tacon},\ and\ \citenamefont {Keimer}}]{Souliou2017}%
  \BibitemOpen
  \bibfield  {author} {\bibinfo {author} {\bibfnamefont {S.-M.}\ \bibnamefont
  {Souliou}}, \bibinfo {author} {\bibfnamefont {J.}~\bibnamefont {Chaloupka}},
  \bibinfo {author} {\bibfnamefont {G.}~\bibnamefont {Khaliullin}}, \bibinfo
  {author} {\bibfnamefont {G.}~\bibnamefont {Ryu}}, \bibinfo {author}
  {\bibfnamefont {A.}~\bibnamefont {Jain}}, \bibinfo {author} {\bibfnamefont
  {B.~J.}\ \bibnamefont {Kim}}, \bibinfo {author} {\bibfnamefont
  {M.}~\bibnamefont {Le~Tacon}},\ and\ \bibinfo {author} {\bibfnamefont
  {B.}~\bibnamefont {Keimer}},\ }\bibfield  {title} {\bibinfo {title} {{Raman
  Scattering from Higgs Mode Oscillations in the Two-Dimensional
  Antiferromagnet ${\mathrm{Ca}}_{2}{\mathrm{RuO}}_{4}$}},\ }\href
  {https://doi.org/10.1103/PhysRevLett.119.067201} {\bibfield  {journal}
  {\bibinfo  {journal} {Phys. Rev. Lett.}\ }\textbf {\bibinfo {volume} {119}},\
  \bibinfo {pages} {067201} (\bibinfo {year} {2017})}\BibitemShut {NoStop}%
\end{thebibliography}%

\end{document}